\documentclass[journal]{IEEEtran}

\usepackage{amsmath,url, amssymb}
\usepackage{mathrsfs}
\usepackage[pdftex]{graphicx}
\usepackage[caption]{subfig}
\usepackage{balance}
\usepackage{enumitem, kantlipsum}
\usepackage[mathscr]{euscript}
 \let\mathscr\relax
\usepackage[scr]{rsfso}
\usepackage{textcomp}
\usepackage{multirow, bm}
\usepackage{makecell}
\usepackage{arydshln}
\usepackage{xcolor}
\usepackage{afterpage}
\usepackage{tikz}
\usepackage{amsfonts}
\usepackage{tabularx}
\usepackage{float}

\def\I{{\mathbf I}}
\def\K{{\mathbf K}}
\def\W{{\mathbf W}}

\def\Kiv{{{\K}_\mathrm{I_V} }}

\def\I{{\mathbf I}}

\def\c{\mathbf c}
\def\Kv{{\mathbf {K}_{\mathrm{V}}}}

\begin{document}

\title{Facing Device Attribution Problem for Stabilized Video Sequences}

\author{Sara~Mandelli,~\IEEEmembership{Student Member,~IEEE,}
        Paolo~Bestagini,~\IEEEmembership{Member,~IEEE,}
        Luisa~Verdoliva,~\IEEEmembership{Senior~Member,~IEEE,}
        and~Stefano~Tubaro,~\IEEEmembership{Senior~Member,~IEEE}
\thanks{S. Mandelli, P. Bestagini and S. Tubaro are with the Dipartimento di Elettronica, Informazione e Bioingegneria, Politecnico di Milano, Milano 20133, Italy (e-mail: name.surname@polimi.it).}
\thanks{L. Verdoliva is with the Dipartimento di Ingegneria Elettrica e delle Tecnologie dell'Informazione, University of Naples Federico II, 80125 Naples, Italy (e-mail: verdoliv@unina.it).}
\thanks{This material is based on research sponsored by DARPA and Air Force Research Laboratory (AFRL) under agreement number FA8750-16-2-0173. The U.S. Government is authorized to reproduce and distribute reprints for Governmental purposes notwithstanding any copyright notation thereon. The views and conclusions contained herein are those of the authors and should not be interpreted as necessarily representing the official policies or endorsements, either expressed or implied, of DARPA and Air Force Research Laboratory (AFRL) or the U.S. Government.}
}

\maketitle

\begin{abstract}
A problem deeply investigated by multimedia forensics researchers is the one of detecting which device has been used to capture a video.
This enables to trace down the owner of a video sequence, which proves extremely helpful to solve copyright infringement cases as well as to fight distribution of illicit material (e.g., underage clips, terroristic threats, etc.).
Currently, the most promising methods to tackle this task exploit unique noise traces left by camera sensors on acquired images.
However, given the recent advancements in motion stabilization of video content, robustness of sensor pattern noise-based techniques are strongly hindered.
Indeed, video stabilization introduces geometric transformations between video frames, thus making camera fingerprint estimation problematic with classical approaches.
In this paper, we deal with the challenging problem of attributing stabilized videos to their recording device.
Specifically, we propose:
(i) a strategy to extract the characteristic fingerprint of a device, starting from either a set of images or stabilized video sequences;
(ii) a strategy to match a stabilized video sequence with a given fingerprint in order to solve the device attribution problem.
The proposed methodology is tested on videos coming from a set of different smartphones, taken from the modern publicly available Vision Dataset.
The conducted experiments also provide an interesting insight on the effect of modern smartphones video stabilization algorithms on specific video frames.
\end{abstract}


\section{Introduction}\label{sec:introduction}

\IEEEPARstart{T}{he} vast majority of traffic flowing over the Internet is composed of visual data, especially videos.
More and more often, videos are used to support news, not only by information professionals, but also by end users of social networks.
Besides their explicit message,
such videos carry a wealth of implicit information which can be exploited for forensic tasks \cite{Milani2012}, first of all source attribution.
Linking a given video to its acquisition device may provide precious evidence both during investigations and before a court of law.
For example, it can allow to expose copyright violations, or point to the authors of hideous crimes, such as acts of terrorism or pedo-pornography.
The key assumption for source identification is that acquisition devices leave distinctive traces in the acquired content.
Therefore, these traces can be exploited to retrieve information on the origin of the video at various levels of granularity, that is, brand, model, or individual device \cite{Kirchner2015}.
This latter information, of course, is the most valuable and sought for.

To date, the most powerful methods for device identification rely on the camera photo-response non uniformity (PRNU) pattern.
The PRNU pattern is due to inhomogeneities in silicon wafers and imperfections of the sensor manufacturing, 
which cause a non-uniform sensitivity to light of the sensor photo-diodes.
As a results, a deterministic multiplicative noise component can be observed in all images or videos acquired by the same camera.
Each device is characterized by its unique PRNU pattern, which can be therefore regarded as a sort of camera fingerprint.
Due to its properties, the PRNU pattern allows reliable device identification \cite{Lukas2006}, even in the presence of JPEG compression \cite{Chen2008}.
Moreover, it can be used for other forensic tasks, such as image forgery detection \cite{Chen2008, Chierchia2014}.
The PRNU-based approach normally relies on some prior information, typically a large number of images known to come from the camera of interest,
however, blind methods have been also proposed with competitive performance \cite{Lin2017, Marra2017}.
Use of compressed PRNU patterns has been also proposed \cite{Valsesia2015, bondi2019} to allow real-time applications.

PRNU-based methods have been readily extended to video to accomplish a variety of forensic tasks,
e.g., source identification \cite{chen2007}, detection of duplicate or spliced videos \cite{Bayram2008, Mandelli2018} authentication of smartphones \cite{Galdi2017}.
This extension, however, is not easy, and several peculiar issues need to be addressed to obtain a satisfactory performance.
In fact, PRNU estimation is much harder for videos than for images,
since videos are almost always compressed with relatively low quality, and often subjected to video stabilization.

Gaining robustness against compression is a primary goal of current research,
since videos are most often uploaded on YouTube \cite{vanHouten2009} or shared through other social networks \cite{Amerini2017, Meji2018}.
In \cite{chen2007} blockiness artifacts caused by compression are corrected before evaluating decision statistics.
In \cite{McCloskey2008} a confidence weighting scheme is proposed to identify high-frequency areas of the scene, which are discarded to ensure a more reliable PRNU estimation.
In \cite{Chuang2011} video frames are reordered and weighed according to their reliability given that I-frames are more reliable than P-frames for estimation.
Also, videos delivered on a wireless network suffer from blocking and blurring due to packet losses,
and suitable algorithms need to be developed to handle this situation \cite{Chen2015}.

Another major problem is video stabilization,
by which individual frames can undergo geometrical transformations (e.g., translation, scale, rotation, etc.) after acquisition to compensate for involuntary user's movements \cite{grundmann2018}.
This causes misalignment of individual pixels across frames, preventing a reliable estimation of the PRNU fingerprint.
In addition, even when it is correctly estimated, it may not correlate with the noise residuals extracted from a given stabilized frame.
Since modern smartphone cameras adopt video stabilization, and most of the videos uploaded on the internet come from smartphones,
PRNU-based methods may be of little use \cite{shullani2017} without suitable corrections.

The first paper addressing this problem \cite{Hoglund2011} dates back to 2011, but it only compensates for translations.
In \cite{taspinar2016} it is more realistically assumed that stabilization is performed using a combination of translation and rotation,
which are estimated and compensated before evaluating correlation using only I-frames.
It is also proposed to perform video camera attribution using a set of images from the same camera
and it has been shown that the fingerprint computed from a set of images correlate with the fingerprint extracted from a non-stabilized video of the same camera.
This idea is further developed in \cite{iuliani2017} where an hybrid sensor pattern noise analysis is carried out
to handle the problem of video stabilization.
Specifically, the reference PRNU is estimated using only still images,
while for the test video some scale and translation transformations are performed to register frames to the image reference.

In this paper, we face the problem of camera attribution from stabilized video sequences exploiting PRNU-based traces.
Specifically, we propose two different approaches to extract the camera fingerprint, using either images or stabilized videos obtained from the same device.
We then propose a methodology to test a video sequence (even if stabilized) against a fingerprint for camera attribution.
A simplified version of this methodology is also proposed for situations in which computational complexity is a constraint, and many video frames are available.
The proposed camera attribution strategy is tested on the publicly available Vision dataset \cite{shullani2017}, consisting of almost 400 stabilized and non-stabilized video sequences obtained from modern portable devices.

In terms of our novel contributions, we would like to highlight the following aspects:
\begin{itemize}
	\item We propose a solution based on modeling video stabilization by means of similarity transformations, thus compensating for scale, rotation and translation operations motivated by \cite{grundmann2018}.
    \item We propose the first method for camera attribution only using stabilized videos, tested in a completely uncontrolled scenario of videos stabilized by proprietary software (i.e., camera firmwares).
    \item We propose a strategy based on a global optimization technique, rather than using a brute force approach for stabilization parameters' estimation as in \cite{taspinar2016, iuliani2017}. This makes the proposed method more suitable for realistic applications.
    \item We discuss the interesting scenario of performing camera attribution when the first frame acquired by the camera is not available (e.g., the video has been trimmed in time). As a matter of fact, the first frame is often non-stabilized, thus making camera attribution simpler but making the scenario less realistic.
\end{itemize}

The rest of the paper is structured as follows.
Section~\ref{sec:intro_method} reports some background on video stabilization and PRNU, and introduces the problem formulation.
Section~\ref{sec:reference} contains the details of the two approaches proposed to estimate a reference fingerprint from the available images and/or videos.
Section~\ref{sec:k_test} explains the proposed algorithm for testing a video query against the previously obtained camera fingerprints.
Section~\ref{sec:results} reports a detailed overview of the performed experimental campaign.
Finally, Section~\ref{sec:conclusions} concludes the paper.

\section{Background and problem statement}
\label{sec:intro_method}

In this section we introduce some background concepts useful to understand the rest of the paper.
First, we overview recently proposed methods for video stabilization.
Then, we introduce the concept of photo response non-uniformity (PRNU) and its use for camera attribution, highlighting the problems that arise when dealing with videos.
Finally, we report details about the formulation of the video camera attribution problem faced in this paper.

\subsection{Video motion stabilization}
\label{sec:intro_stab}
Because of the indiscriminate sharing of video content on social media platforms, the amount of video sequences posted on the web is increasing every day.
Since a significant percentage of these videos is captured by amateur users, which are usually not equipped by professional stabilization tools (e.g., tripods, steady-cam, etc.), the recorded sequences may highly suffer from camera-shake, principally due to the hand-held recording but also to other movements induced by the user, walking or even running during capture. 

As a consequence, plenty of strategies to perform the stabilization of a video (directly on the recording camera or off-line) have been proposed \cite{Gleicher2008, liu2009, Karpenko2011, Ringaby2012, grundmann2012, grundmann2018}.
These methods allow to improve the quality of the recorded videos, making each sequence appearing as if it were recorded from a stable camera moving along a smooth path.
In particular, these systems are able to detect and correct high frequency jitter artifacts, low frequency artifacts, rolling shutter wobbles, foreground motion, poor lighting, and scene cuts \cite{grundmann2018}. 

Among the most recently proposed approaches, the authors of \cite{grundmann2018} propose to perform video stabilization by fitting the original 2D camera path with linear motion models, characterized by a different amount of degrees of freedom (DOF).
Whenever these models are considered to be valid for the considered frame-pair motion, the original path is transformed according to the model and a smooth camera path is generated.
Frames are then warped on this new path by applying a set of pixel-wise transformations.

To be precise, the authors show that it is possible to increase the complexity of the algorithm by considering motion models with more DOF.
The easiest motion model describes only translations, hence 2 DOF. 
Specifically, the motion of each frame can be represented as function of that of the previous one by means of the matrix $\mathbf{T}_2$, defined as
\begin{equation}
\mathbf{T}_2 = \begin{pmatrix}
1  & 0 &c_x\\ 
0 & 1 &c_y
\end{pmatrix} , 
\label{eq:2_dof}
\end{equation}
where $c_x$ and $c_y$ are the magnitude of translation of the camera along the horizontal and vertical axes, respectively.
Alternatively, the similarity model including 4 DOF to detect also rotation and uniform scaling between frames can be used.
The matrix describing the motion relationship between frames is
\begin{equation}
\mathbf{T}_4 = \begin{pmatrix}
s \cdot \cos \theta & -s \cdot \sin \theta &c_x\\ 
s \cdot \sin \theta& s \cdot \cos \theta &c_y
\end{pmatrix} ,
\label{eq:4_dof}
\end{equation}
being $s$ and $\theta$ the scaling factor and rotation angle, respectively. 
More complex homographic models can be also considered if perspective distortions have to be recovered.
However, not every model can efficiently represent the motion between two frames, and the application of an incorrect motion model introduces distortions in the stabilized video.
As an example, whenever the model is invalid, translation and similarity inject additional shaking in the estimated path, whereas the homographic models result in perspective warping errors. 
Moreover, the higher the complexity of the used model, the higher the probability of wrongly estimating it, potentially leading to temporal instability of the generated path \cite{liu2009, grundmann2011}.

In the light of this, stabilization methods usually perform a first step to delete the shake due to similarity and lower DOF motions, without taking into account higher DOF motions.
Then, any residual motion can be potentially corrected exploiting the homographic models \cite{grundmann2018}.
This two-step approach comes in handy whenever computational complexity is an issue.
Indeed, if stabilization is performed on mobile devices, even just one step can be used.

A consequence of motion stabilization on a video sequence is that two pixels sharing the same geometrical coordinates on two different frames may have been acquired with different portions of the camera sensor due to the introduced geometrical transformations.
As shall be clear in the next section, this is a problem for PRNU-based video camera attribution.

\subsection{Photo response non-uniformity}
The PRNU is a noise fingerprint characteristic of any image and video acquisition device. 
Specifically, PRNU is introduced in all acquired images and video frames as a multiplicative zero-mean noise pattern \cite{Lukas2006, Chen2008}. 


In the basic procedure proposed in the literature \cite{Chen2008},  PRNU is estimated from a set of $ N $ images $ \I_n$ coming from the same device as
\begin{equation}
\K = \frac{\sum\limits_{n = 1}^{N} {\W}_n \cdot \I_n  }{ \sum\limits_{n = 1}^{N}  \I_n^2}  \, , 
\label{eq:prnu_def}
\end{equation}
where $ \W_n $ is the noise residual extracted from $ \I_n$, and all operations are performed pixel-wise. Precisely, $ \W_n= \I_n- \dot{\I}_n $, being $ \dot{\I}_n $ a denoised version of $ \I_n $ computed as suggested in \cite{Chen2008}.

The PRNU $ \K $ can be exploited as a camera signature for solving the image-camera attribution problem, i.e., given a test image $ \I $, inferring whether it has been captured by the camera or not. 
For instance, one way to solve the problem is based on measuring Peak-to-Correlation-Energy (PCE) \cite{chen2007}.
More precisely, the PCE defines a measure of the cross-correlation between 2D data arrays.
In order to compute it over two matrices $ \W_1 $ and $ \W_2 $ of size $ m \times n $, we first have to cross-correlate them, defining $ [\mathbf{R}]_{u,v}= \mathrm{corr}([\W_1]_{i,j}, [\W_2]_{i-u,j-v})$. Then, by examining the $ \mathbf{R}$ surface it is possible to detect the presence of a pronounced peak.
Thus, the $ \mathrm{PCE}(\W_1, \W_2) $ is defined as
\begin{equation}
\mathrm{PCE}(\W_1, \W_2) =  \frac{[\mathbf{R}^2]_{u_p,v_p}   }{ \frac{1}{m \cdot n - |\mathcal{N}_p| }    \cdot \sum\limits_{u, v \notin \mathcal{N}_{p}} [\mathbf{R}^2]_{u,v}      }      \, , 
\label{eq:pce_def}
\end{equation}
where $ (u_p, v_p) $ are the peak coordinates and $ \mathcal{N}_p $ is a small neighborhood of the peak \cite{chen2007}.
In other words, the PCE measures how much two matrices are correlated, independently of possible shift misalignments between them.
Indeed, $ (u_p, v_p) $ can be seen as an estimate of the mutual shift between $\W_1$ and $\W_2$.

Dealing with the image-camera attribution problem, we can compute the PCE between the noise residual $ \W $ extracted from the test image and the camera PRNU pixel-wise scaled by $ \I $, denoted as $\mathrm{PCE}(\W, \K \cdot \I )$. If this value is higher than a confidence threshold, $\I $ is attributed to that camera \cite{Lukas2006, Chen2008}. 
Indeed, the PCE is a very good metrics for solving the problem, as it is robust by nature to shifts.
This is very important whenever an image has been cropped with respect to the reference PRNU \cite{goljan2008}.
By means of PCE, the peak of correlation can be searched over all the pixel positions, hence possible errors due to shifts between PRNU and the extracted noise are avoided.

Extending PRNU-based methods to video sequences is not straightforward, and presents multiple challenges \cite{chen2007, taspinar2016}. 
In fact, video signals are typically less reliable than image ones due to their lower resolution, as well as stronger compression.
Therefore, PRNU traces in video sequences tend to be very subtle.
As a matter of fact, the authors of \cite{chen2007,Chuang2011} propose to consider each video frame as a picture and follow the standard PRNU-based pipeline for image attribution.
Their results confirm that video attribution is challenging, and not all video frames can be considered as equally informative.
Indeed, depending on the used coding strategy, intra-coded frames typically contain more reliable PRNU information.

Interesting alternative approaches have been proposed by \cite{taspinar2016, iuliani2017}.
The authors suggest to estimate each camera reference PRNU from images, and use it to attribute video queries.
However, to solve the camera attribution problem for a video query they cannot directly make use of the PRNU reported in \eqref{eq:prnu_def}, as the video resolution is typically lower than the resolution of the images taken by the same camera \cite{chen2007}.
As a matter of fact, in order to adapt the sensor size to the video recording area, some operations are performed.
Therefore, they propose a strategy that searches for a correct scale and crop transformation to match image PRNU and video resolutions. 

\begin{figure}[t]
	\centering	\includegraphics[width=\columnwidth]{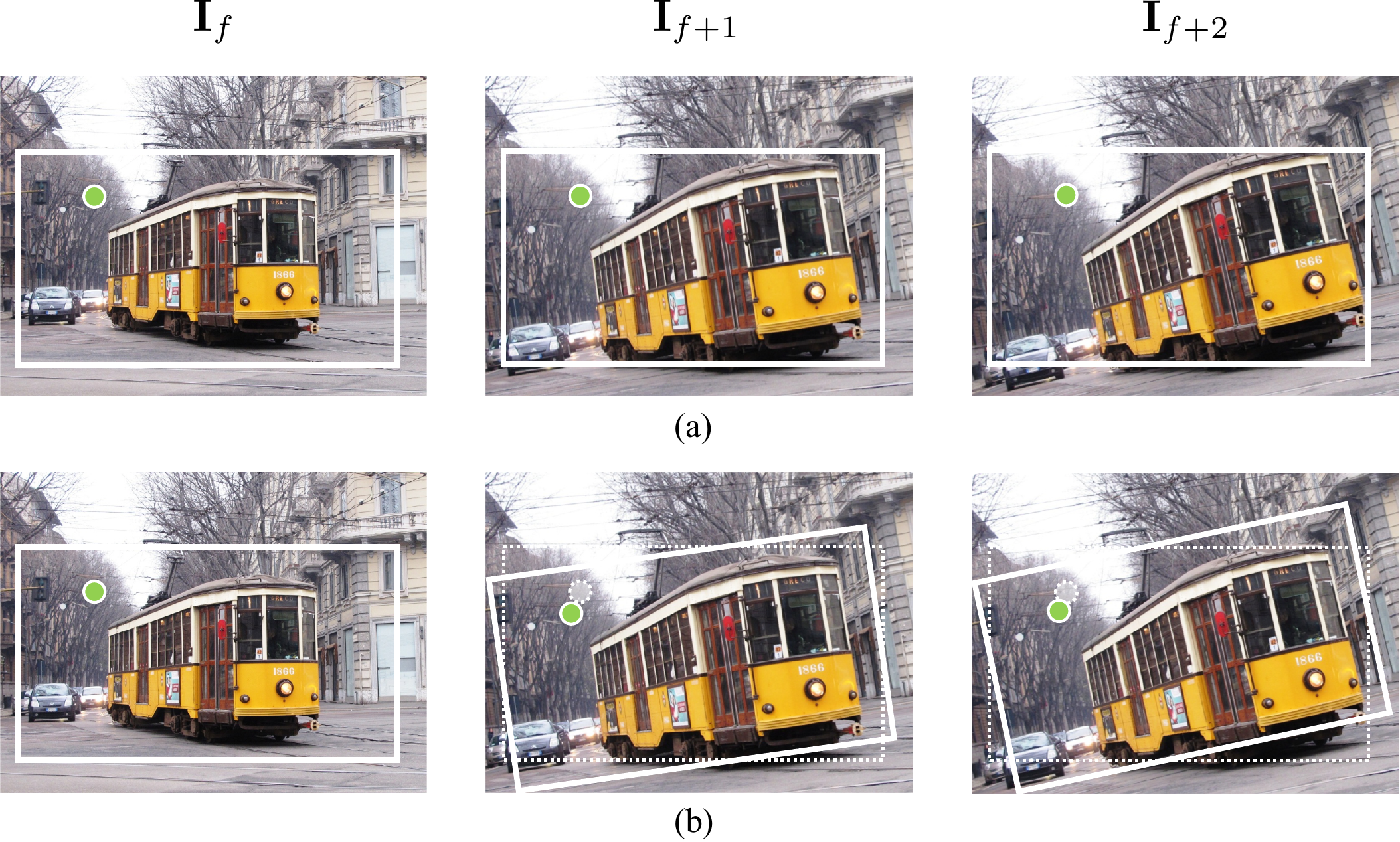}
	\caption{(a) In absence of stabilization, the pixel coordinates (green circle) do not modify on frames; (b) In presence of stabilization, recording sensor area can be slightly shifted, scaled and rotated, hence the pixel coordinates (green circle) can change with respect to the original ones (shown in dashed line).}
	\label{fig:stab_nostab}
\end{figure}

Though, if video stabilization is used, additional transformations are carried out as well, as reported in Section \ref{sec:intro_stab}.
In the light of these consideration, we propose to exploit a 4-parameter linear model to describe the image-to-video space conversion, modeling both the operation that shrinks the recording area and the stabilization counteracting global frame shake \cite{liu2009, grundmann2012}.

The considered linear model consists in a 2D similarity transformation $ \mathscr{T} $ \cite{grundmann2011}, described by the matrix $ \mathbf{T} $ as
\begin{equation}
\mathbf{T} = \begin{pmatrix}
s \cdot \cos \theta & -s \cdot \sin \theta &c_x\\ 
s \cdot \sin \theta& s \cdot \cos \theta &c_y
\end{pmatrix} . 
\label{eq:similarity}
\end{equation}
The transformation of image to video domain can be modeled as $ \mathcal{I}_f = \mathscr{T} (\mathcal{I}) $, being $ \mathcal{I} $ the image space and $ \mathcal{I}_f $ the space related to video frames. 
In particular, the vector $ \c = (c_x, c_y )$ describes the translation along the $x$ and $y$ axes, whereas $ s $  and $ \theta $ regard scale and rotation, respectively. 

Concerning non-stabilized videos, the relationship expressed by \eqref{eq:similarity} can be further simplified by noticing that no kind of rotation is reasonably included in video acquisition process.
As a matter of fact, there would be no reason to rotate the sensor area whenever a non-stabilized video is recorded.
Moreover, as reported in \cite{iuliani2017}, all video frames recorded by a unique non-stabilized device are affected by equal scaling and shift factors, being these parameters probably fixed by the device firmware specifications. 

On the contrary, whenever video stabilization is used, each frame $\I_f$ in the sequence experiences its own scale $ s_f $, translation $ \c_f $ and rotation $ \theta_f $, which are introduced to drop the shaky-hand effect typical of amateur recorded videos.
Therefore, all these transformations (i.e., scale, rotation and shift) should be taken into account when dealing with PRNU-related problems.

For the sake of clarity, Fig.~\ref{fig:stab_nostab} reports three adjacent frames of a video. The area inside the white window points out the final scene depicted on the recorded video by the device. In the former row, the depicted scene in absence of stabilization: selecting a pixel inside the recording area, its coordinates maintain fixed during capture. Whenever stabilization is present (latter row), in order to generate a stable camera path, each pixel can actually vary its coordinates during the recording.

\subsection{Problem formulation}

In this paper, we focus on the problem of video source attribution exploiting PRNU-based traces in the challenging scenario of in-camera stabilized video sequences.

In order to solve this problem, we split it into two separate steps:
(i) given some multimedia content acquired with a device, estimate its PRNU-based fingerprint.
(ii) given this fingerprint and a video query, detect whether the video comes from the camera under analysis.

Concerning the first step, the primary goal is finding a good estimation of the camera fingerprint in the video resolution domain.
To this purpose, it is reasonable to consider three main strategies, depending on the data owned by the analyst:
(i) exploiting only images shot by the camera to estimate $ \K $, then transform it into video domain, given that the conversion parameters reported in \eqref{eq:similarity} are known;
(ii) exploiting both images and videos shot by the camera, without knowing the conversion parameters;
(iii) exploiting only videos recorded by the camera.
To the best of our knowledge, the first case is not realistic as the warping image-to-video parameters are not apriori known neither reported in the literature.
Therefore, we focus on the other two scenarios.

Once the video fingerprint has been estimated, we need to compare it to video queries.
Therefore, we propose a solution (and a simplified version of it) to accomplish this task.

\begin{figure}[t]
	\centering	\includegraphics[width=\columnwidth]{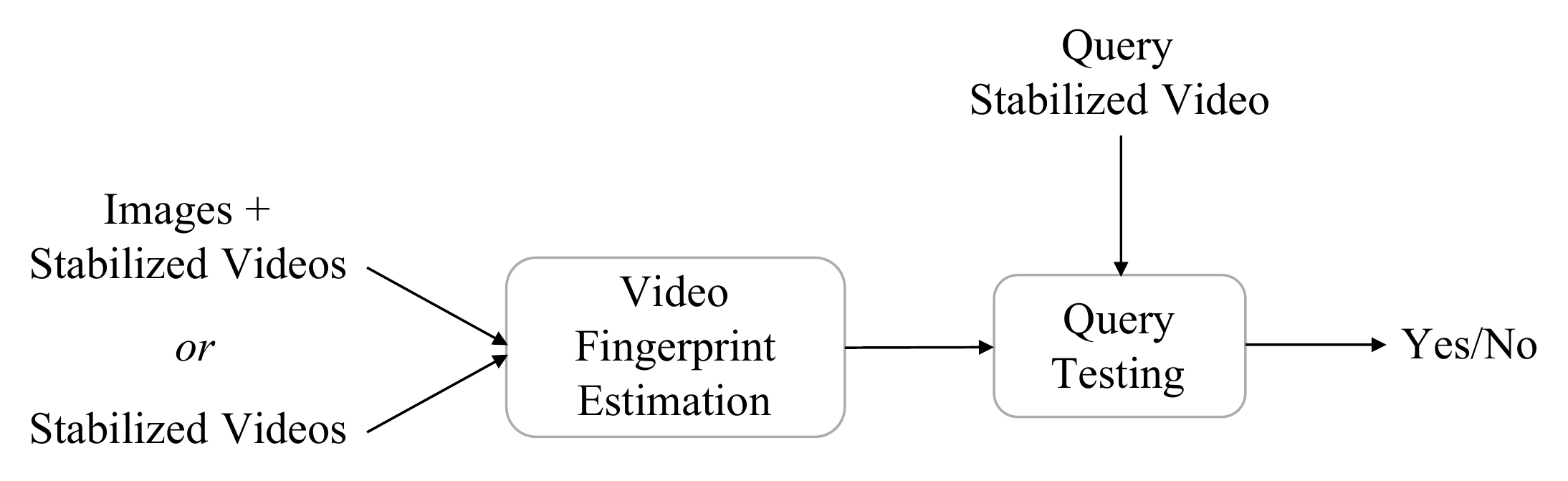}
	\caption{Pipeline of the proposed method: Initially, the device signature is estimated, using images and videos captured by the camera, or videos only. Then, each video query is tested and eventually attributed or not to the camera. }
	\label{fig:pipeline}
\end{figure}

Specifically, Fig.~\ref{fig:pipeline} depicts the pipeline of the proposed method. 
In the following, we present the proposed strategies for fingerprint estimation and video source attribution, discussing on the main intuitions behind the approaches.

\section{Reference video fingerprint estimation}\label{sec:reference}

In this section we explain how to estimate the reference video fingerprint to characterize a video camcorder.
This is the camera signature which a video should be compared with to solve attribution problems.
Specifically, we consider two different working scenario.
In the first one, the analyst can use some images as well as videos coming from the same device.
In the second one, the analyst has only a set of videos to be used to characterize the camera.
In the following we report all the details for both scenarios.

\subsection{Reference video fingerprint from images and videos}
\label{sec:ki_v_est}

The first video fingerprint estimation scenario we consider is that of an analyst who has a set of images and videos recorded with the same camera.
In this setup, we propose a pipeline composed by three main steps:
\begin{itemize}
	\item Estimate the device image PRNU $ \K $ from the available set of pictures applying \eqref{eq:prnu_def}.
    \item Estimate the image-to-video transformation parameters $s$, $\theta$, and $\mathbf{c}=(c_x, c_y)$ used in \eqref{eq:similarity} by solving an iterative maximization problem.
    \item Estimate the device video fingerprint $ \Kiv $ by warping $ \K $ with the estimated parameters $s$, $\theta$, and $\mathbf{c}$.
\end{itemize}
In other words, we propose to use as video fingerprint a transformed image PRNU downsampled to the scale of video frames resolution.
This choice is driven by the following observations:
\begin{itemize}
	\item Images are often acquired at higher resolution and with better coding quality than videos, thus typically containing more reliable device fingerprint information.
    \item Downsampling image PRNU to the video frames scale requires much less computational power than upsampling video frames to the scale of the image PRNU (e.g., smaller matrices to fit into memory, PCE correlation is computed on less samples, etc.).
    \item It has been shown in \cite{bondi2019} that PRNU downscaling of a factor up to $2$ does not hinder significantly camera attribution performance, which is good news considering that image resolution is not often twice that of a video.
\end{itemize}


In order to estimate the parameters $s$, $\theta$, and $\mathbf{c}$ of the selected device, we search for the geometric transformation that maximizes the PCE correlation between the transformed version of $\K$ and frames residuals $\W_f$,  extracted from frames $\I_f$ belonging to a set $\mathcal{F}$ of selected frames.
To be precise, we select reasonable search ranges $ \mathcal{S}, \mathcal{T} $, $ \mathcal{C} $ for scale, rotation angle and shift, respectively.
Then, we estimate $ s_f, \theta_f, \c_f$ for each frame by solving the maximization problem
\begin{equation}
s_f, \theta_f, \c_f= \underset{s \in \mathcal{S}, \theta \in \mathcal{T}, \c \in \mathcal{C}}{\arg \max} \,  \mathrm{PCE}(\W_f, \; \I_f \cdot \mathscr{T}_{s_f \theta_f \c_f}(\K)). 
\label{eq: max_pce_img_vid}
\end{equation}
This maximization problem is solved using iterative methods.
In particular, we use a particle swarm optimization technique \cite{Kennedy2011}.

If frames come from a non-stabilized video sequence, $s_f$, $\theta_f$ and $\c_f$ are expected to be coherent for all frames.
Indeed, cameras typically do not use different portions of the sensor from frame to frame.
Moreover, $\theta_f$ is expected to be zero, as typically non-stabilized videos are not acquired rotating the sensor.
This is confirmed by \cite{taspinar2016, iuliani2017}, in which the authors only seek for scale and translations, but not rotation.
Therefore, any tuple $s_f$, $\theta_f=0$ and $\c_f$ can be used as $s$, $\theta$, and $\mathbf{c}$ estimate.

If frames come from \textit{stabilized} video sequences, $s_f$, $\theta_f$ and $\c_f$ can vary from frame to frame, as each frame is independently warped depending on the content to stabilize.
However, if the considered video does not contain strongly textured areas (flat scene are always suggested for PRNU extraction \cite{Chen2008}) and it is not characterized by excessive device shaking (typically true for videos to be pleasant at visual inspection), we can assume that $s_f$, $\theta_f$ and $\c_f$ only slightly change from frame to frame, oscillating around the true values $s$, $\theta$ and $\c$. 
Furthermore, it is reasonable to assume that the rotation contribution is likely to be very small, and it is almost zero for at least a small set of frames.
Indeed, the captured scene should not look rotated to video viewers.

Therefore, in order to select a unique parameter set for the image-to-video PRNU conversion, we propose to fix $\theta=0$, and average the estimated $s_f$ and $\c_f$ parameters over the frames with strong PCE.
Notice that frames for which rotation parameters is not really zero can be filtered out from our estimate as they will be characterized by low PCE value.

Formally, we compute 
\begin{equation}
p_f= \mathrm{PCE}(\W_f, \; \I_f \cdot \mathscr{T}_{s_f \theta_f \c_f}(\K)).
\label{eq: max_pce}
\end{equation}
Then, we estimate scale and translation parameters as
\begin{equation}
s =  \sum_{f \in \mathcal{F}_\mathrm{I}} \frac{s_f}{|\mathcal{F}_\mathrm{I}|}, \;\;\;
c_x=  \sum_{f \in \mathcal{F}_\mathrm{I}} \frac{c_{x_f }}{|\mathcal{F}_\mathrm{I}|}, \;\;\;
c_y =  \sum_{f \in \mathcal{F}_\mathrm{I}} \frac{c_{y_f}}{|\mathcal{F}_\mathrm{I}|},
\end{equation}
being $\mathcal{F}_\mathrm{I}$ the set of frames for which $p_f > 60$ (i.e., a PCE threshold suggested in \cite{entrieri2016, taspinar2016}) and  $|\mathcal{F}_\mathrm{I}|$ its cardinality.

\begin{figure}[t]
	\centering
	\includegraphics[width=.45\columnwidth]{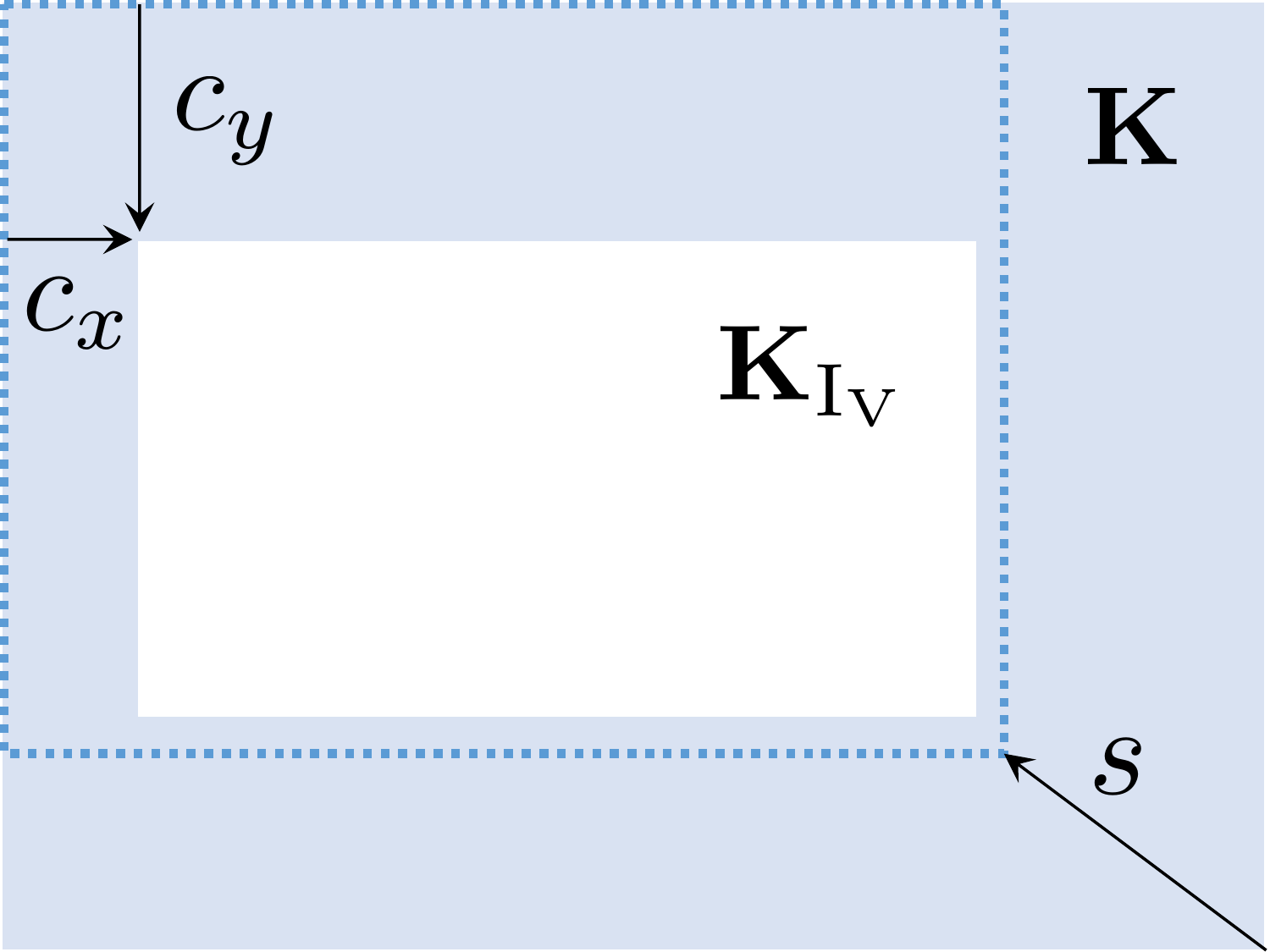}
	\caption{Image PRNU conversion to the video domain. Blue area is the whole sensor size, used for image PRNU, whereas white area represents the video fingerprint $\Kiv$. }
	\label{fig:i_to_v_conv}
\end{figure}

In order to pass from image to video domain, we apply the similarity transformation $ \mathscr{T}_{s \c}(\cdot)$ to the image PRNU, thus obtaining the video fingerprint
\begin{equation}
	\Kiv = \mathscr{T}_{s \c}(\K).
\end{equation}
For the sake of clarity, Fig.~\ref{fig:i_to_v_conv}  depicts the operations done for converting the image PRNU to the video domain.
A scale transformation with parameter $s$ is performed on $\K$ to shrink the space to a reduced area, then scene is cropped to match video resolution, according to the estimated shift $\c$. 

Note that $\Kiv$ can be exploited as device signature for testing the camera attribution problem over a generic video query.
Indeed, the resolution of the fingerprint should approximately match the resolution of the query sequence. 
Moreover, since $\Kiv$ is the result of aligned noise contributions coming from high resolution images, it reasonably contains a very reliable camera model information.


\subsection{Reference video fingerprint from videos only}
\label{sec:kv_est}

The second video fingerprint estimation scenario we consider is that of an analyst who only has a set of videos recorded with the same camera.
Indeed, information about images captured by the device under analysis is not always available. 
For this reason, we focus on how to estimate the device fingerprint directly from video content, even in the challenging scenario of motion stabilization.
In this setup, we propose a pipeline composed by the following steps:
\begin{itemize}
	\item We search for the frame whose residual $\W_f$ correlates well in terms of PCE with the largest number of other frames' residuals.
    \item We estimate a candidate video fingerprint $\Kv$ starting from the selected $\W_f$.
    \item We update the video fingerprint $\Kv$ by iteratively aggregating information from other frames whose noise residuals correlate well with the fingerprint in terms of PCE.
\end{itemize}

Despite this pipeline seems trivial, the procedure of noise aggregation is not straightforward at all.
Indeed, due to motion stabilization, frames are shifted, rotated and scaled one another.
Hence, if we randomly pick a set of frames and estimate the fingerprint following PRNU estimation as reported in \eqref{eq:prnu_def}, noise residuals left by the camera sensor risk to be averaged while misaligned, eventually contributing with very few content of the original fingerprint.

To avoid averaging misaligned contributions, noise residuals should be in principle coherently warped one on the other by following a procedure similar to the one proposed in Section~\ref{sec:ki_v_est}.
However, sensor noise traces are extremely subtle in video signals due to the typically used aggressive lossy coding schemes.
To top it all off, scene content often leaks into frame noise residuals due to the used suboptimal denoising algorithms.
These two factors make the estimation of transformation parameters that map a frame noise residual on another one an almost preposterous task.

In order to avoid mistakenly estimating warping parameters still being able to estimate a reliable video fingerprint, we make the assumption that a set of video frames affected by (almost) the same stabilization transformation exists within the available reference videos.
This assumption reasonably holds for sequences characterized by low (if any) textured content that does not need to be too much stabilized (i.e., typical sequences used for PRNU estimation).
Under this assumption, we propose an iterative noise residual aggregation method composed by the following steps:
\begin{enumerate}[label=\textit{\alph*)}, wide, labelwidth=!, labelindent=8pt]
\item Loop over all the frames in the set $\mathcal{F}$ of available ones.
For every frame $\I_f, \, f \in \mathcal{F} $, solve the standard camera attribution problem for the other frames with respect to noise residual $ \W_f $. 
Specifically, compute $\mathrm{PCE}(\W_l, \I_l \cdot  \W_f), \forall l, f \in \mathcal{F}, \, l \ne f$.


\item Analyze the relative PCE values in search for frames whose residuals match.
A match is considered if two constraints on PCE are satisfied.

The first constraint is on PCE magnitude.
We consider a match only for strictly positive PCE values to avoid strongly uncorrelated frames.

As second constraint, we check the relative shift between frames (i.e., the position of PCE maximum value).
As a matter of fact, the effect of video stabilization is to scale, rotate and translate the frames one another, but without introducing visible artifacts on the recorded sequence.
For this reason, it is reasonable to assume that a stabilization algorithm does not translate too much one frame with respect to the temporally adjacent ones. 

In principle, if the frames were not stabilized, the relative shift estimated through PCE should be of $ (0, 0)$ pixels, since both frames residuals should be aligned in terms of sensor noise.
Conversely, in stabilized videos, the relative alignment can be different from $ (0,0)$.
However, under the hypothesis of small translations introduced by stabilization, if the relative alignment is too far from $ (0,0)$ we can attribute it mainly to PCE correlating textured content or additional noise contributions, rather than noise patterns related to the original camera fingerprint.

Therefore, in order to avoid false matching results that do not actually correspond to noise residuals' alignment, we only consider matching residuals if the relative shift is of $ (\Delta, \Delta)$ pixels ($ \Delta = \{5, 10, 20, 30\}$ in our experiments with Full-HD sequences).

\item Select as reference frame $\I_r$ the video frame $ \I_f $ that matches with the largest number of frames according to the matching definition provided in step (b).
The first video fingerprint candidate is the residual of frame $ \I_r $, namely  $ \Kv = \W_r$. 
Include $\I_r$ to the set of frames exploited for the fingerprint estimation, defined as $ \mathcal{F}_\mathrm{V} $.

\item Correlate the remaining frames with the estimated fingerprint, computing $\mathrm{PCE} (\W_f, \I_f \cdot \Kv)$.

\item Add frames that honor the constraints reported in (b) to the set $ \mathcal{F}_\mathrm{V} $ of frames useful to estimate the fingerprint, after compensating for the relative shift misalignment with respect to the estimated fingerprint.


\item Update the estimated fingerprint $ \Kv $ by averaging all the noise residuals $ \W_f, f \in \mathcal{F}_\mathrm{V}  $.


\end{enumerate}
Iterate steps (d), (e), (f) until no more frame residuals matching with $ \Kv $ are left.
Eventually, the estimated camera fingerprint for testing the video queries is $ \Kv $.

Notice that, the big difference between $ \Kiv $ (estimated following the procedure reported in Section~\ref{sec:ki_v_est}) and $ \Kv $ (estimated following the reported in Section~\ref{sec:kv_est}) is that the former is an aggregation of image noise residuals, whereas the latter is an aggregation of video noise residuals.
For this reason, $ \Kiv $ can be considered a higher quality estimate of the device video fingerprint compared to $ \Kv $.

\section{testing the video query}
\label{sec:k_test}
Once we estimate the reference fingerprint for each camera, namely either $ \Kiv $ or $ \Kv $, we can proceed in solving the camera attribution problem for every video query.

Given a generic camera fingerprint $\K $ and a video to be attributed, we propose to test each frame of the sequence following a similar procedure to the standard PCE-based method.
Specifically, since video stabilization introduces different geometric transformations from frame to frame, we estimate the warping configuration which maximizes the PCE between each transformed frame and the fingerprint $ \K $.
In this way, even if the fingerprint has been already registered into video domain, we can compensate for the additional stabilization deviations introduced on every frame of the video query.

Therefore, exploiting a particle swarming optimizer (PSO) \cite{Kennedy2011}, we estimate the scaling, the rotation angle and the relative shift for every frame in the sequence such that the PCE is maximized, i.e.,
\begin{equation}
P_f =\underset{s \in \mathcal{S}, \theta \in \mathcal{T}, \c \in \mathcal{C}}{\max} \, \mathrm{PCE}(\mathscr{T}_{s \theta \c}(\W_f), \mathscr{T}_{s \theta \c}(\I_f) \cdot  \K),
\label{eq:pce_test_warp}
\end{equation}
where $f$ is the frame index belonging to the set $\mathcal{F}$ of considered query frames, with cardinality $F$.

Then, in order to attribute or not the query video to the camera, we simply select the highest $ P_f $ over all tested frames as
\begin{equation}
P_{q}=\underset{f \in \mathcal{F}}{\max} \, P_f . 
\end{equation}
If $P_{q}$ overcomes a certain threshold, the query is attributed to the camera, otherwise it is considered coming from a different device.

This approach empirically proves to be quite accurate, and we shall show with our experimental campaign that even a reduced number of frames is enough for performing a correct video query matching.
However, this methodology is time consuming, as the estimation of the warping parameters through particle swarming optimization requires a fairly high amount of operations.

In order to overcome this obstacle, we propose one possible way out, which can be very efficient whenever there is a consistent amount $F$ of query frames.
As a matter of fact, it is likely that not all frames in the set underwent strong rotation or scale transformations due to stabilization.
As reported in \cite{grundmann2018}, it is common to exploit reduced motion model including translation only to stabilize some video frames, at the benefit of faster estimation and greater stability. 
Hence, we can limit our search to the estimation of the relative shift between the query frames and the fingerprint, still likely finding some frames with high PCE. 
To test the whole video, we select again the best PCE obtained over the set as
\begin{equation}
P_{q}=\underset{f \in \mathcal{F}}{\max} \, \mathrm{PCE}(\W_f, \I_f \cdot  \K).
\end{equation}
In doing so, we automatically discard from the test all query frames that actually underwent rotation or scaling, as they will provide low PCE.
To attribute the query video to the device under analysis, we threshold $P_{q}$.

Concerning both proposed methods, robustness strongly depend on the length of the query video.
Intuitively, the larger the frame-set, the higher the probability to find one correlating frame over the whole video query.
We shall show in the next section how these strategies represent viable solutions for solving the camera attribution problem in presence of video stabilization.

\section{Results}\label{sec:results}
In this section we report the results of the conducted experimental analysis. 
First, we describe the used dataset, then we define the adopted evaluation metrics, and finally we report the numerical results achieved by the proposed method. 
In doing so, we also discuss other state-of-the-art methods for camera attribution using stabilized video sequences.

\subsection{Dataset}\label{subsec:dataset}
In order to test our method in a fair setup, we make use of a dataset of almost 400 videos coming from 24 different devices.
This dataset has been built starting from the recently released Vision Dataset, which includes images and videos from a wide variety of mobile devices from $ 11 $ major brands \cite{shullani2017}.

In particular, for building the image PRNU $\K$ of each device, we select all the available images shot by the device depicting scenes of flat surfaces, as smooth images are always suggested for PRNU estimation \cite{Chen2008}.

Concerning the video dataset, we select all videos with resolution equal to Full-HD ($ 1920 \times 1080 $).
For each device, we consider both static and motion scenes (corresponding to the tags \emph{still, panrot, move} in \cite{shullani2017}).
Moreover, we also include videos with almost-flat content and with a significant texture contribution (i.e., labeled as \emph{flat, indoor, outdoor} in \cite{shullani2017}).
In doing so, we end up with 165 non-stabilized sequences from $10$ devices, and 232 stabilized video sequences from $14$ devices.
For each sequence, with an average temporal duration of one minute, we only make use of I-frames, as they contain more reliable sensor noise information with respect to inter-predicted frames \cite{Chuang2011, taspinar2016}.

When we refer to any specific device, we make use of the same naming convention introduced in \cite{shullani2017}.

\subsection{Evaluation Metrics}\label{subsec:metrics}
In order to assess the accuracy in solving camera attribution problem, we resort to receiver operating characteristic (ROC) curves.
Specifically, for each camera, we consider all the videos recorded with that camera as positive samples, whereas the set of negatives includes an equal number of sequences not taken with that camera, randomly selected from the dataset.
Each curve depicts the resulting relationship between true positive rate ($ \mathrm{TPR} $) and false positive rate ($ \mathrm{FPR} $), averaged over the set of available cameras. 
To numerically evaluate the quality of the attribution, we use these parameters:
\begin{itemize}
	\item $ \mathrm{AUC} $, defined as the area under the (ROC) curve, the higher the better.
	\item $ \mathrm{TPR}_{@0.01} $, defined as the $ \mathrm{TPR} $ calculated ad a fixed $ \mathrm{FPR} $ of $ 1 \% $. 
\end{itemize}
The goal is reaching a high value of $ \mathrm{AUC} $ (ideally 1) and the highest possible value for $ \mathrm{TPR}_{@0.01} $ as well.

\subsection{Preliminary study on stabilization disadvantages}\label{subsubsec:visionas}
In order to confirm the difficulty of dealing with stabilized video sequences, we perform a preliminary test by facing camera attribution problem using standard procedures devised for non-stabilized videos.

\begin{figure}[t]
	\centering	\includegraphics[width=.65\columnwidth]{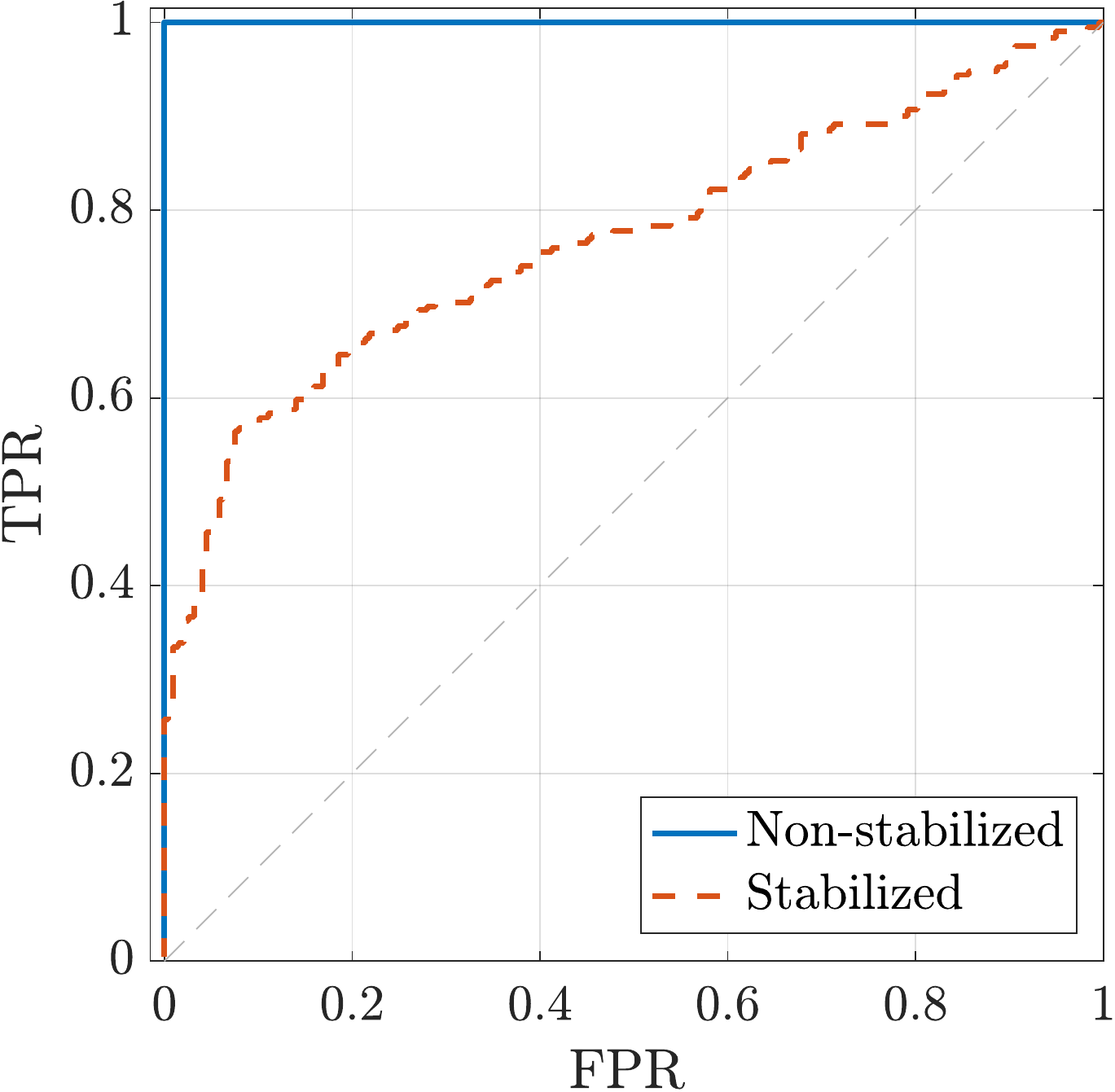}
	\caption{ROC curves obtained using the standard PRNU-based video source attribution method \cite{chen2007} considering non-stabilized and stabilized sequences. Video stabilization strongly hinders the state-of-the-art performances.}
	\label{fig:ROC_visionas}
\end{figure}

Similarly to approaches proposed in \cite{chen2007, shullani2017}, we compute the fingerprint of each video by simply aggregating noise residuals extracted from I-frames.
The reference camera fingerprint is estimated selecting a static and low-textured sequence for each device, precisely the first video tagged as \emph{flat}-\emph{still} in the dataset. 
For testing a generic query, we compute the PCE between the camera fingerprint and the query one.
We apply the same pipeline to both non-stabilized and stabilized video sequences.

Results are depicted in Fig.~\ref{fig:ROC_visionas}.
The difference between stabilized and non-stabilized videos ROC curves is evident.
For the non-stabilized pool the pipeline achieves $ \mathrm{AUC} = 1 $, which means perfect device attribution.
For the stabilized set this pipeline achieves much lower performances with $ \mathrm{AUC} = 0.77 $.
This confirms that video stabilization makes PRNU-based video camera attribution a more challenging task as shown in \cite{taspinar2016,iuliani2017}.

\subsection{Considerations about the first video frame}\label{subsubsec:1fr_res}
The authors of \cite{grundmann2018} show that, generally, the first frame of a video does not undergo any stabilization, as there is no motion to be corrected. 
Indeed, it is possible that the first frame is taken as reference for stabilizing the next frames.
On one hand, this is good news whenever the first video frame is available to the analyst.
On the other hand, we must realistically assume that the video sequence under analysis might have been temporally trimmed, thus making the first acquired video frame unavailable.

In order to test the effect of using or not the first frame for camera attribution with stabilized videos, we perform the following experiment.
We estimate the reference fingerprint $ \Kiv $ for each device.
Then, we proceed with the testing phase, following both strategies reported in Section~\ref{sec:k_test}, considering a single frame for each query video (i.e., $F=1$).
The test has been performed in two scenarios: i) selecting the first frame (i.e., $f=1$); ii) selecting a random I-frame different from the first one (i..e, $ f \neq 1 $). 

\begin{figure}[t]
	\centering
	\includegraphics[width=.65\columnwidth]{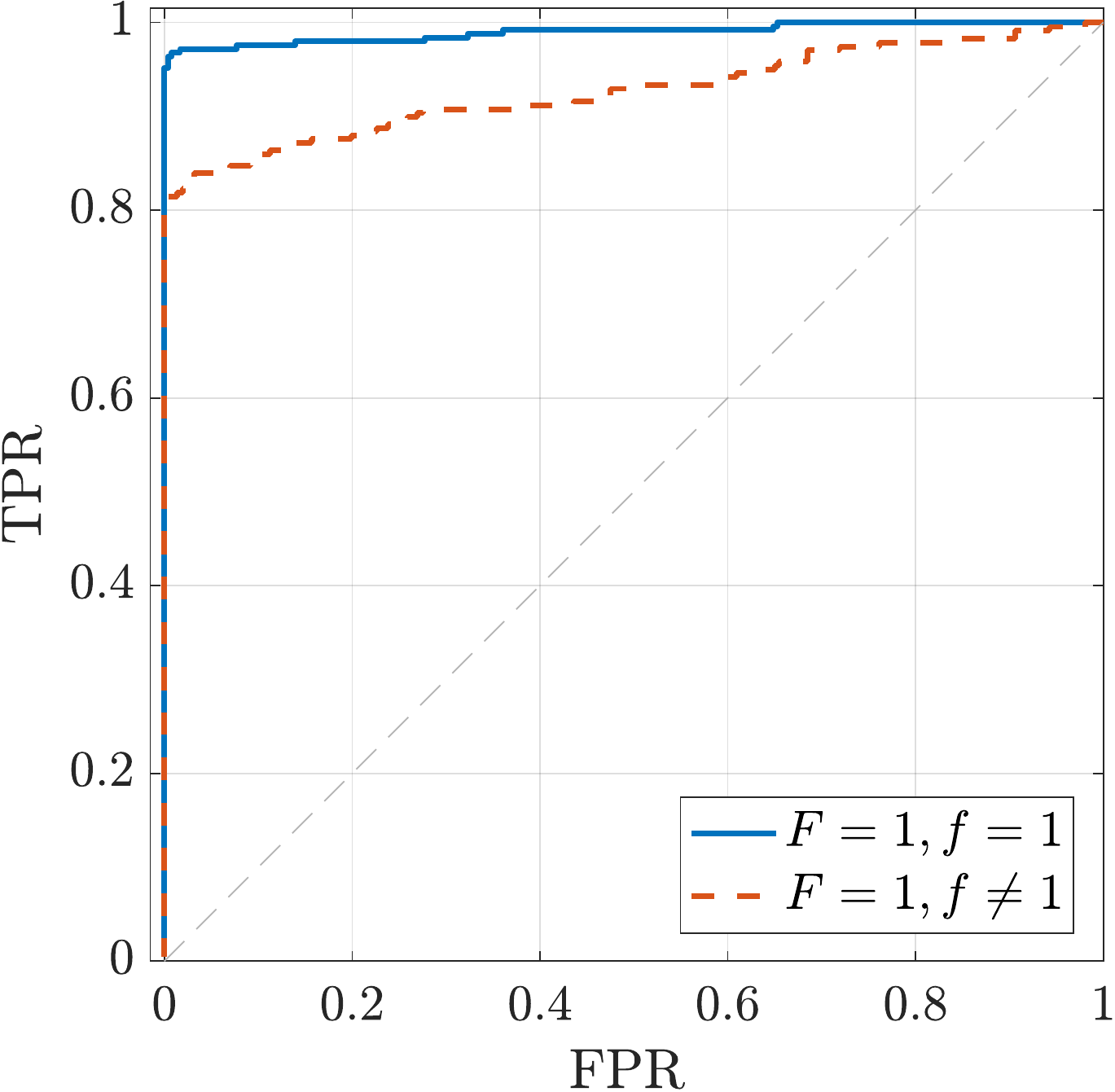}
	\caption{ROC curve related to a single-frame video, exploiting the complete test strategy.}
	\label{fig:ROC_ki_1fr_warp}
\end{figure}

\begin{figure}[t]
	\centering
	\includegraphics[width=.65\columnwidth]{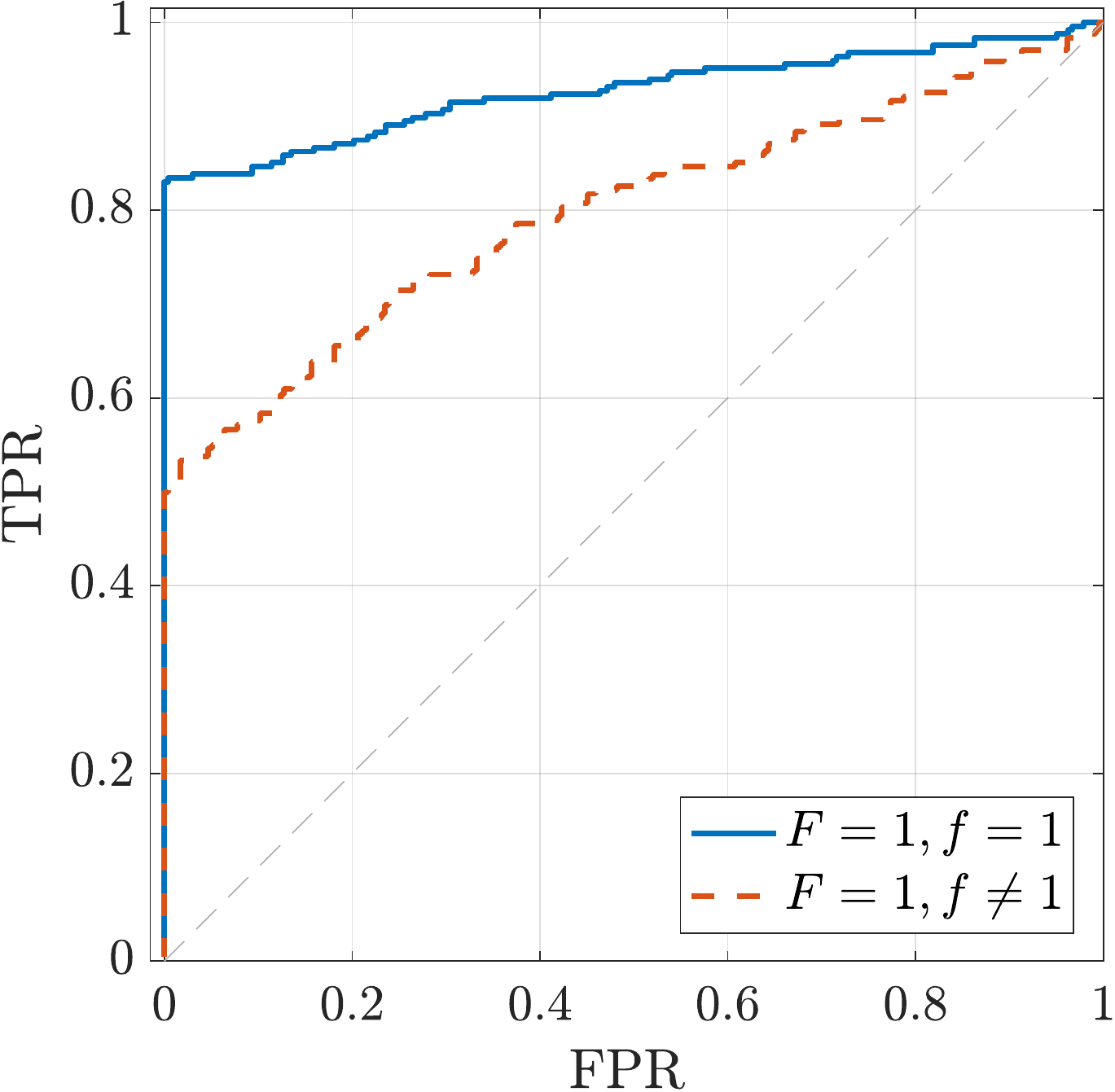}
	\caption{ROC curve related to a single-frame video, exploiting the quick test strategy.}
	\label{fig:ROC_ki_nowarp_1fr}
\end{figure}

ROC curves for the complete test strategy that registers query frames on the fingerprint are reported in Fig.~\ref{fig:ROC_ki_1fr_warp}.
ROC curves for the quick test strategy that only searches for translations are reported in Fig.~\ref{fig:ROC_ki_nowarp_1fr}.
In both situations, results obtained considering the first frame are well above the ones computed with a random frame.
This confirms that stabilization algorithms used by devices within Vision dataset may skip stabilization on the first frame.

In order to avoid biasing the results and making wrong conclusions about the proposed algorithm, we never include the first frame in our experiments, neither for the estimation of the video camera fingerprint nor for the testing phase.
In doing so, we assume working in the far more general scenario in which videos available at the analyst can be short portions of longer sequences.

\subsection{Reference video fingerprint from images and videos}\label{subsubsec:alg_Ki_res}
To estimate the video fingerprint $ \Kiv $, we need to estimate some scaling and translation parameters for each device as explained in Section~\ref{sec:ki_v_est}.
To estimate the image to video domain transformation, we apply the proposed $ \Kiv $ estimation algorithm to $ 10 $ randomly selected I-frames coming from reasonably flat and almost static scenes from each stabilized device in the dataset.
In particular, we select as reasonable search range for the scaling factor $ \mathcal{S} = [0.3, 0.85] $, whereas the shift is searched all over the video resolution.

Table~\ref{tab:ki_to_hki} reports the estimated average scale (i.e., $s$) and translation (i.e., $c_x$ and $c_y$) parameters for each stabilized device, following the same nomenclature used in \cite{shullani2017}.
These parameters are used to obtain a $ \Kiv $ estimate for each device, starting from the PRNU $\K$ computed from images.

\begin{table*}[t]
	\caption{Average scaling and shift parameters for image-to-video domain conversion. Device naming convention is the same as in \protect\cite{shullani2017}. Only stabilized devices have been used.}
	\label{tab:ki_to_hki}
	\centering
	\def\arraystretch{1.8}
	\resizebox{\textwidth}{!}{
		\begin{tabular}{c|c|c|c|c|c|c|c|c|c|c|c|c|c|c}
			DEVICE		& D02   & D05     & D06     &D10		& D12     & D14 & D15   & D18  &  D19  & D20  & D25  & D29    & D32  & D34          \\  \Xhline{1.5pt}
			$ {s } $      & $ 0.75 $  & $ 0.687 $  & $ 0.707 $ & $ 0.75 $  &  $ 0.379 $   & $ 0.688 $   & $ 0.706 $   & $ 0.688 $   & $ 0.706 $  & $ 0.815 $  & $ 0.517 $  & $ 0.687 $ & $ 0.517 $   & $ 0.687 $\\
			${ c_x}  $   & $ 270   $   & $ 158 $      & $ 201 $    & $ 279  $    &  $ 33  $  & $ 167  $     & $ 190 $   & $ 166 $  &  $ 190 $  & $ 97 $    &  $ 239 $   & $ 161 $  & $ 242  $   & $ 161   $           \\
			$ {c_y}  $   & $ 374 $     & $ 304 $      & $ 328 $   & $ 384 $     & $ 205 $   & $ 308 $    & $ 323 $  & $ 308  $  &  $ 324 $ & $ 248 $ & $ 361 $    &  $ 302 $ & $ 356 $   & $ 302   $            \\
		\end{tabular}}
	\end{table*}

\subsection{Reference video fingerprint from videos only}\label{subsubsec:alg_Kv_res}
In order to estimate the camera fingerprint $ \Kv $ from stabilized videos only, we follow the iterative noise aggregation method proposed in Section~\ref{sec:kv_est}.
In particular, for each stabilized device, we select an almost static video with little image content (precisely, the first video flagged as \emph{flat}-\emph{still} in the dataset), considering different values for the parameter $ \Delta $ driving the shift search range.
To evaluate whether the aggregation method is correctly working independently from the query test algorithm, we start from the following idea.
We expect $ \Kv $ estimate to become better and better as long as we aggregate correctly more frames.
Conversely, $ \Kv $ estimate should worsen if we aggregate frames without correctly compensating for motion stabilization effects.

In the light of this, we define $ \rho(f) $ as the normalized cross correlation (NCC) between the fingerprint $ \Kiv$ and the fingerprint $ \Kv(f) $, namely the estimated $\Kv$ fingerprint by the aggregation of $ f $ frames.
In order to compare these terms, we estimate the similarity transformation which registers the camera fingerprint $ \Kv $ on $ \Kiv$, then we apply this transformation to the frame-variant $ \Kv(f) $. 

Actually, $ \rho(f) $ can be very helpful to evaluate the algorithm performances in estimating the video fingerprint. 
For instance, if $ \rho(f) $ has a monotonic increasing behavior, we are aligning in a correct way the frames one another.
Otherwise, we are assembling frames by means of some correlating content which has few in common with the original device fingerprint.
The choice of NCC as metrics instead of the PCE is motivated by its higher computational efficiency.
Moreover, being normalized at $ 1 $, NCC allows a clearer comparison between the performance of different devices.

\begin{figure}[t]
	\centering
	\includegraphics[width=.8\columnwidth]{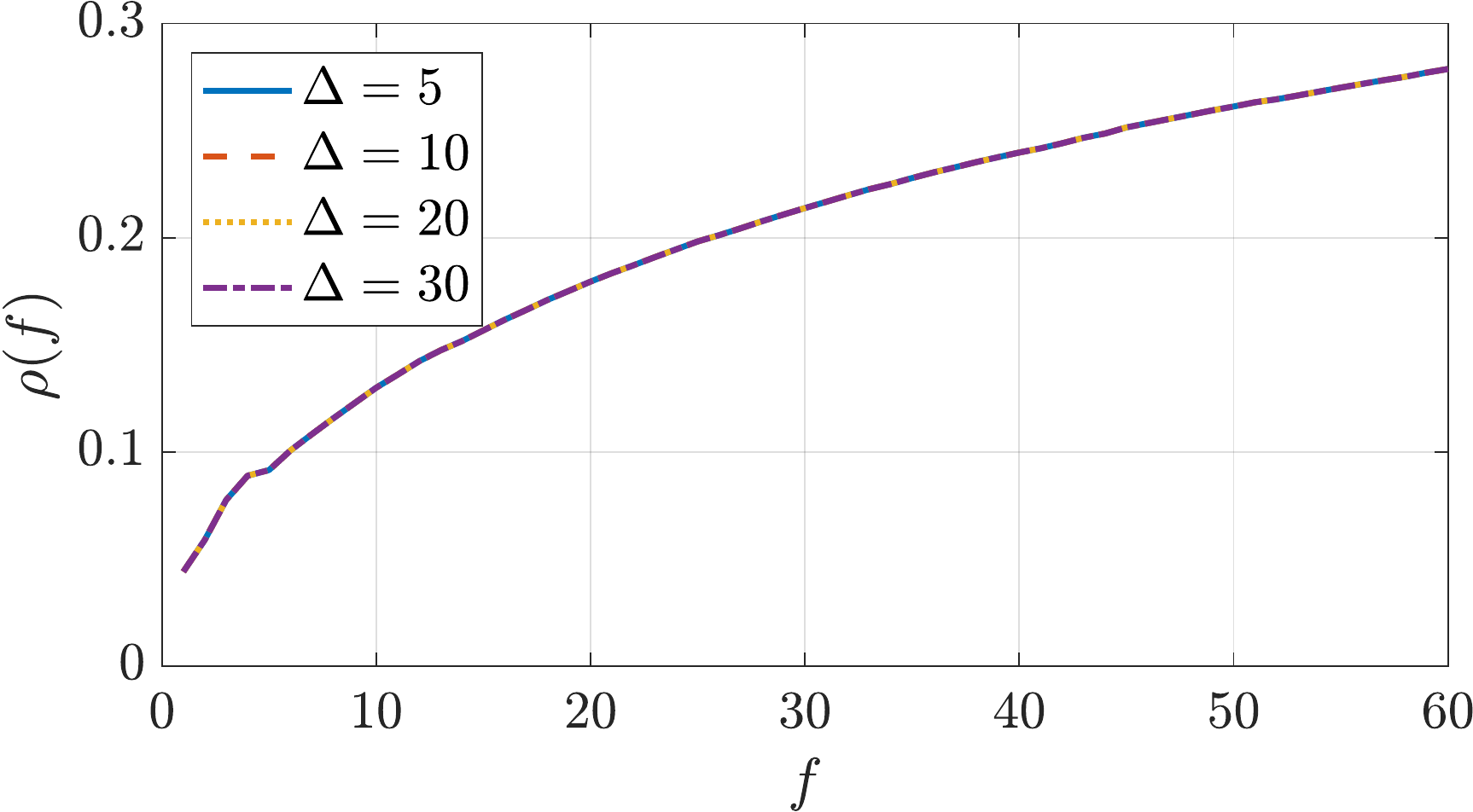}
	\caption{Resulting cumulative NCC computed between $ \Kiv $ and registered $ \Kv(f) $ for device D34. }
	\label{fig:NCC_delta_kv_ki_D34}
\end{figure}

\begin{figure}[t]
	\centering
	\includegraphics[width=.8\columnwidth]{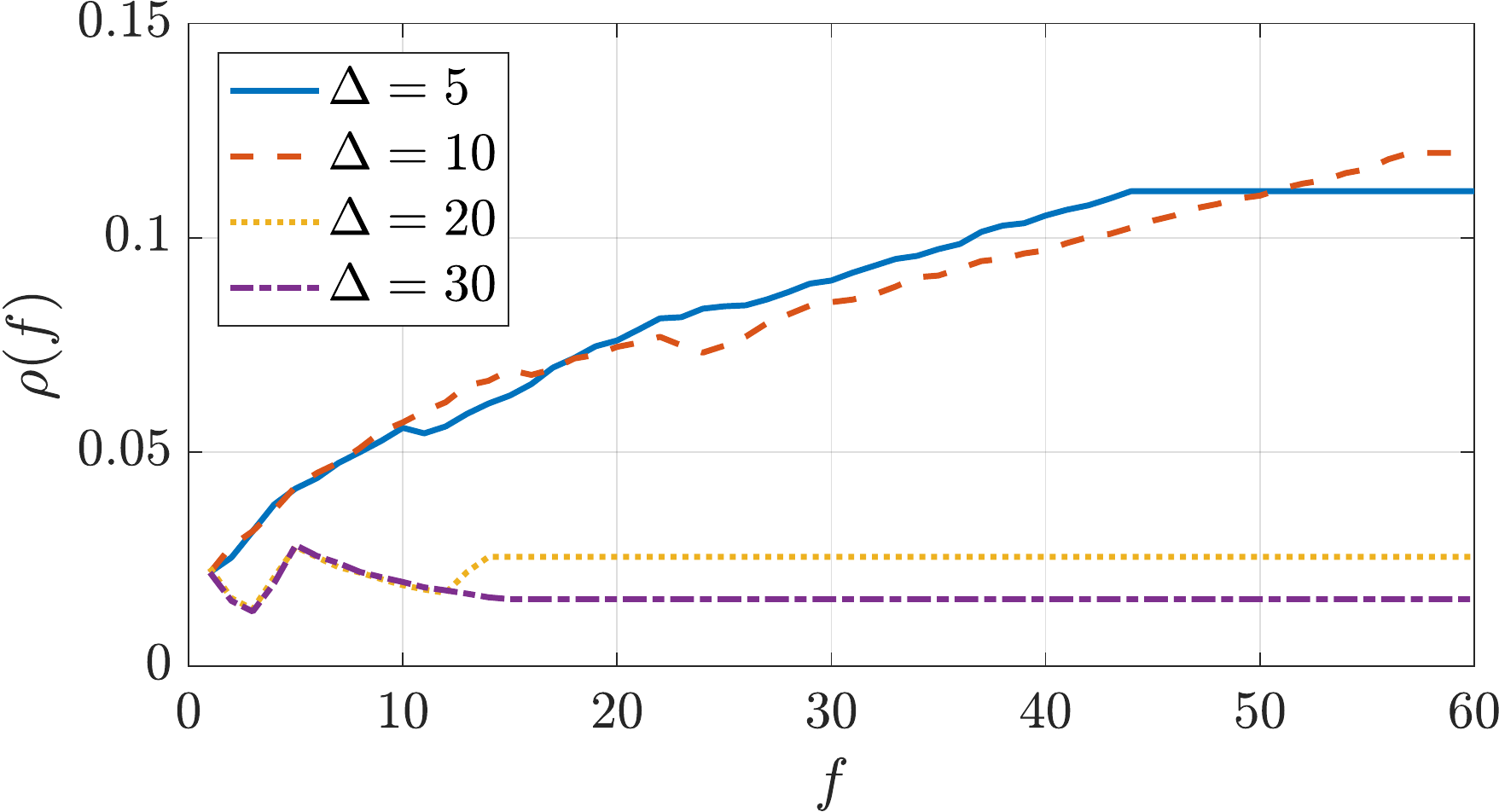}
	\caption{Resulting cumulative NCC computed between $ \Kiv $ and registered $ \Kv(f) $ for device D32.}
	\label{fig:NCC_delta_kv_ki_D32}
\end{figure}

We compute $ \rho(f) $ for each stabilized device in the dataset.
Specifically, results for devices D32 and D34 are reported in Figs.~\ref{fig:NCC_delta_kv_ki_D34} and~\ref{fig:NCC_delta_kv_ki_D32} as representative of the overall trend on all videos.
For some video sequences (see Fig.~\ref{fig:NCC_delta_kv_ki_D34}) the bound on $ \Delta $ value does not impact on the algorithm.
Indeed, $ \rho(f) $ is always monotonically increasing.
However, if videos contain more textures (see Fig.~\ref{fig:NCC_delta_kv_ki_D32}), the proposed registration method tends to register scene content rather than sensor noise traces if $ \Delta $ is too high.
Indeed, $ \rho(f) $ does not increase.
For this reason, we limit our further analysis to $ \Delta = \{5, 10\} $, as these values enable reaching the best aggregation performances on average.

\begin{figure}[t!]
	\centering
	\includegraphics[width=.8\columnwidth]{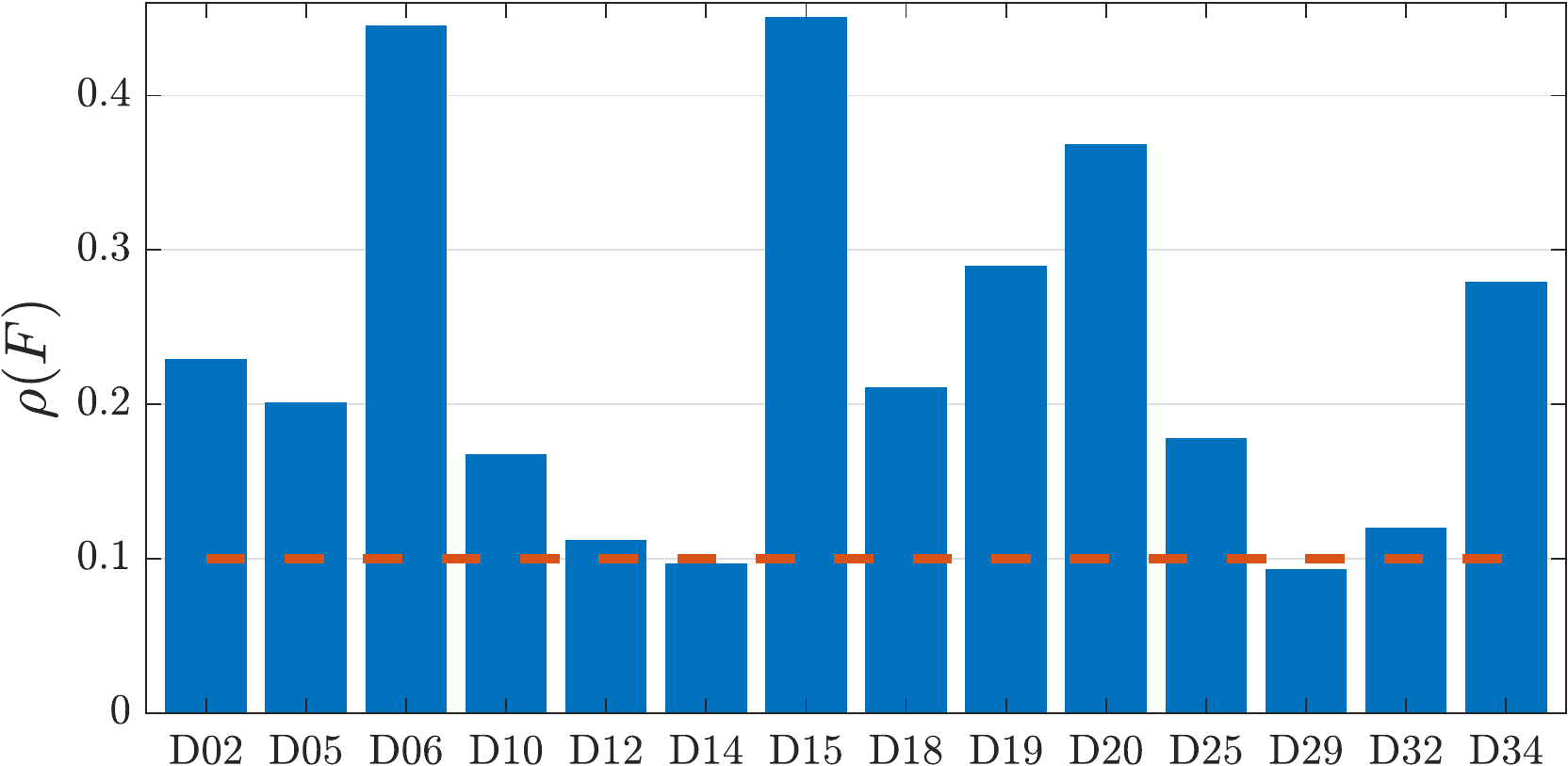}
	\caption{Final value of the resulting cumulative NCC for all stabilized devices in the dataset.}
	\label{fig:NCC_end}
\end{figure}

Finally, Fig.~\ref{fig:NCC_end} shows the values achieved by $ {\rho}(f) $ at the last aggregated frame, namely $\rho(F)$, obtained by fully running the proposed $ \Kv $ estimation method setting $ \Delta = 10 $ for all devices.
Notice that $ {\rho}(F) $ values are almost always greater than $0.1$, which actually represents a good NCC measure in standard attribution problems \cite{Lukas2006}.
This confirms that the estimated video fingerprint $ \Kv $ is informative of the camera model. 

\subsection{Testing the video query}\label{subsubsec:query_res}
To check how accurately we can attribute a stabilized video to the originating device, we present the ROC curves computed over all the stabilized devices in the dataset. 

\begin{table*}[t]
	\caption{$ \mathrm{AUC} $ and $ \mathrm{TPR}_{@0.01} $ exploiting $ \Kiv $ and $ \Kv $ as reference fingerprint, testing $ F =\{5, 10\}$ random I-frames with the complete strategy. }
	\label{tab:ki_kv_kvi_warp}
	\centering
	\def\arraystretch{1.8}
	\resizebox{0.95\textwidth}{!}{
		\begin{tabular}{c|c|c|c|c|c|c}
			$\mathrm{FINGERPRINT}$	  & $ \Kiv, F = 5 $   & $ \Kiv, F = 10 $   & $ \mathbf{K}_{\mathrm{V}_{\Delta = 5}}, F = 5 $     &  $ \mathbf{K}_{\mathrm{V}_{\Delta = 5}}, F = 10$      & $\mathbf{K}_{\mathrm{V}_{\Delta = 10}}, F = 5 $  	& $\mathbf{K}_{\mathrm{V}_{\Delta = 10}}, F = 10 $ 	\\  \Xhline{1.5pt}
			$ \mathrm{AUC} $      				& $ 0.96 $				   & $0.97$	& $ 0.9$                                              & $ 0.89$                     & $ 0.89 $   & $ 0.92 $\\
			$ \mathrm{TPR}_{@0.01} $      & 	$ 0.87 $			& $0.91$					& $  0.71$                                          & $ 0.77 $                       & $ 0.7 $      & $ 0.73 $\\
		\end{tabular}}
	\end{table*}
    
To be precise, for each video sequence, we randomly select $ F $ I-frames, and we test both complete and quick attribution methods on both fingerprints $ \Kiv$ and $ \Kv $. 

First, we report the results achieved using the complete method reported in Section~\ref{sec:k_test}. 
In particular, we set the PSO search range to $ \mathcal{S} = [0.99, 1.01] $ and $ \mathcal{T} = [-0.15, 0.15] $ [rad], following an approach similar to \cite{grundmann2018}.
Indeed, reference fingerprints  $ \Kiv$ and $ \Kv $ are already in the video domain, thus we only need to slightly warp frames for attribution purpose.

\begin{figure}
	\centering
	\includegraphics[width=.65\columnwidth]{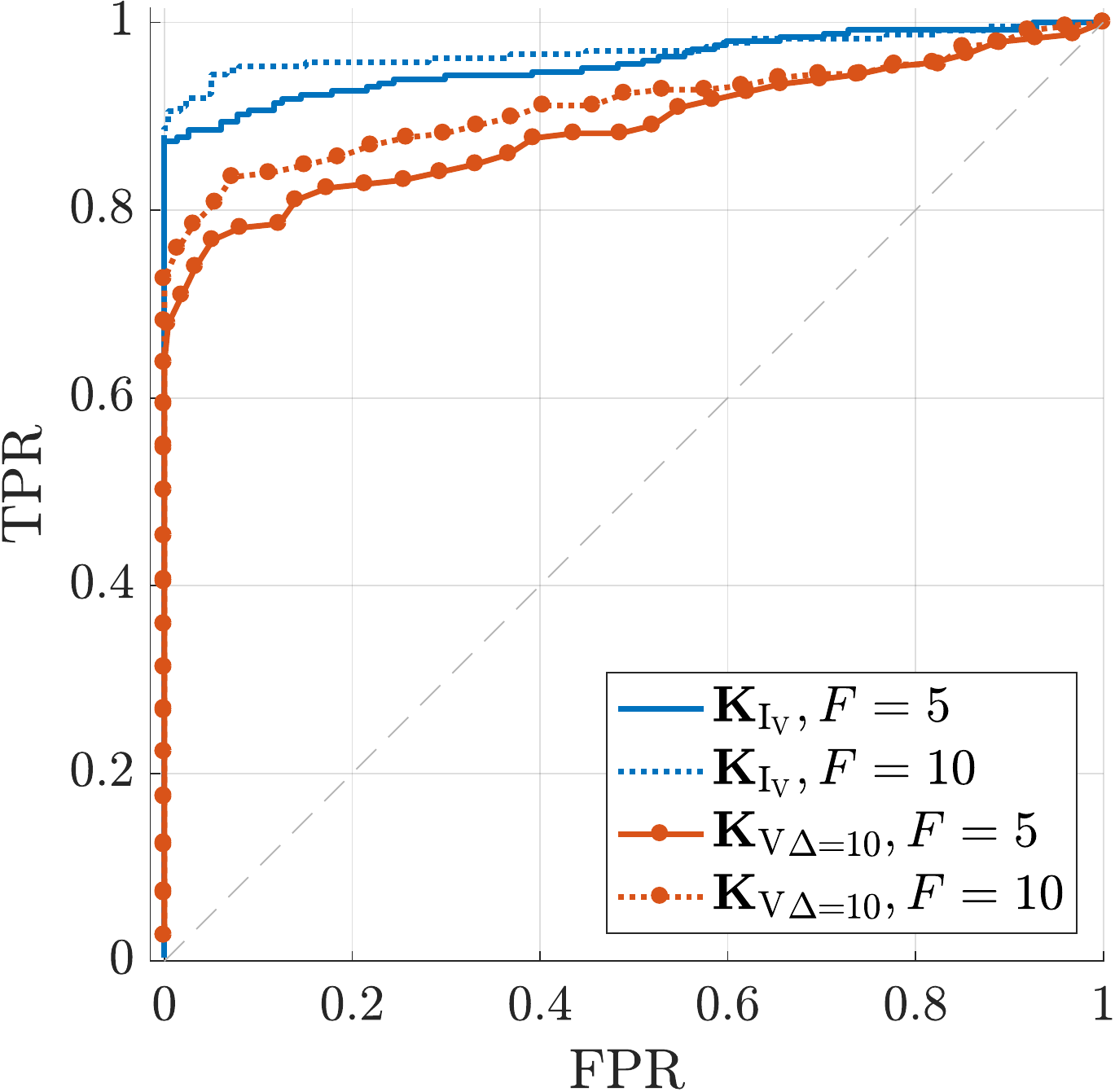}
	\caption{ROC curves obtained testing $ F = \{5, 10\} $ I-frames with the complete strategy.}
	\label{fig:ROC_warp}
\end{figure}

Fig.~\ref{fig:ROC_warp} depicts the results evaluated using both $ \Kiv $ and $\Kv$ as reference fingerprints, and testing $F = \{5, 10\}$ random I-frames of the query videos.
Notice that we only show results for $\mathbf{K}_{\mathrm{V}_{\Delta = 10}}$ ($\Kv$ computed with $\Delta = 10$) as these are highly comparable to the case $\Delta = 5$. 
It is possible to note that proposed method is quite accurate.
As a matter of fact, exploiting just $ 5 $ I-frames (i.e., $ \sim 5 $ seconds of video content), we obtain $ \mathrm{AUC} = 0.96 $ exploiting the fingerprint $\Kiv$. 
On the other hand, performances achieved by $\Kv$ are quite good as well, considering the fingerprint is computed from video frames only. Nonetheless, the larger the amount of investigated I-frames, the better the ROC curve. 
For instance, regarding $\mathbf{K}_{\mathrm{V}_{\Delta = 10}}$ results, with just $ 5 $ frames we can achieve $ \mathrm{AUC} = 0.89 $, whereas $10 $ frames return $ \mathrm{AUC} = 0.92 $.

The overall results of the method are depicted in Table~\ref{tab:ki_kv_kvi_warp}, which reports the achieved $ \mathrm{AUC} $ and $ \mathrm{TPR}_{@0.01} $ corresponding to all the curves. 

In addition, we show the results obtained with the alternative fast attribution method proposed in Section~\ref{sec:k_test}, which is very helpful whether the available computational power is limited and there is a consistent number of query frames. 

\begin{figure}[t]
	\centering
	\includegraphics[width=.65\columnwidth]{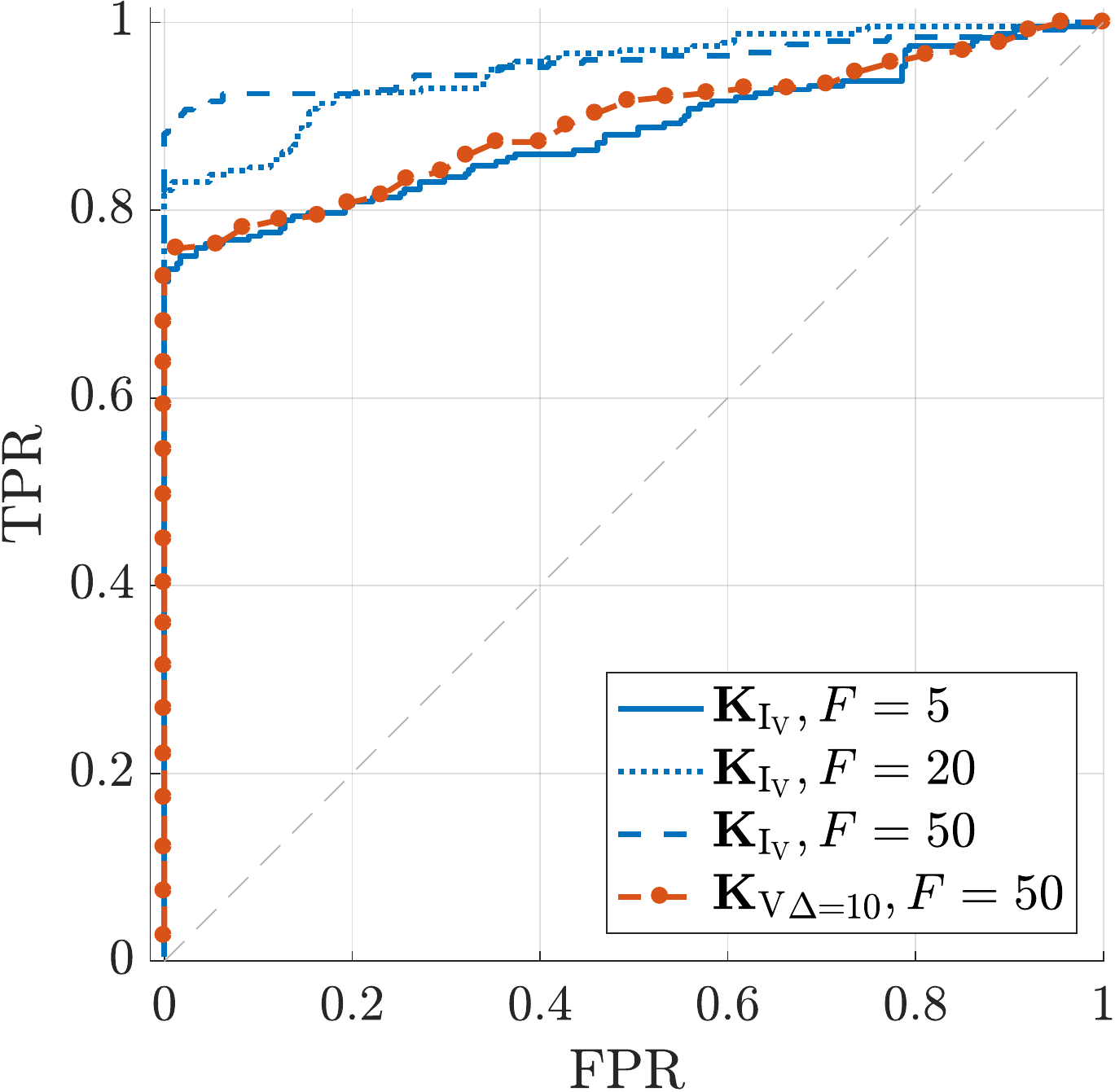}
	\caption{ROC curves obtained testing $ F = \{5, 20, 50\} $ I-frames with the quick strategy.}
	\label{fig:ROC_nowarp}
\end{figure}

Fig.~\ref{fig:ROC_nowarp} depicts the results evaluated using both the reference fingerprints. 
Specifically, we select $F = \{5, 20, 50\} $ query I-frames for evaluating performances of the fingerprint $\Kiv$, whereas for the fingerprint $\Kv$ we directly limit the plots to the use of $50$ I-frames, as a lower amount of frames reduces the performance.
Note that some sequences in the dataset do not have $ 50 $ I-frames. In these situations we use as many I-frames as possible.
We can notice that, to obtain results as accurate as with the complete testing procedure, $F=50$ I-frames are needed rather than just $ 5 $. 
Anyway, it is possible to notice that even exploiting $ \Kv $ as reference we can solve the attribution problem using the fast algorithm, as long as the analyst has approximately one minute long video.


To be precise, Table~\ref{tab:ki_kv_kvi_nowarp} contains $ \mathrm{AUC} $ and $ \mathrm{TPR}_{@0.01} $ corresponding to the curves achieved testing $50$ query I-frames with both reference fingerprints $\Kiv$ and $\Kv$.


\begin{table}[t]
	\caption{$ \mathrm{AUC} $ and $ \mathrm{TPR}_{@0.01} $ exploiting $ \Kiv$ and $ \Kv$ as reference fingerprint, testing $ 50 $ random I-frames following the quick test strategy. }
	\label{tab:ki_kv_kvi_nowarp}
	\centering
	\def\arraystretch{1.8}
	\resizebox{0.8\columnwidth}{!}{
		\begin{tabular}{c|c|c|c}
			$\mathrm{FINGERPRINT}$	                        				& $ \Kiv $			   & $\mathbf{K}_{\mathrm{V}_{\Delta = 5}} $    &  $\mathbf{K}_{\mathrm{V}_{\Delta = 10}} $     		\\  \Xhline{1.5pt}
			$ \mathrm{AUC} $      				& $ 0.96 $			   & $ 0.88 $                      & $ 0.89 $                       \\
			 $ \mathrm{TPR}_{@0.01} $      & 	$ 0.89 $			    & $ 0.74 $                      & $ 0.75 $                    \\
		\end{tabular}}
	\end{table}

As far as the comparison between the complete and fast methods is concerned, we can notice that the former method is more accurate than the latter one as expected.
As a matter of fact, exploiting just $ 5 $ I-frames returns similar results to the case in absence of rotation and scaling only for $ F = 50 $. 
For this reason, when few video frames are available, we suggest to estimate the similarity transformation between the camera signature and the frames.
Indeed, accuracy benefits at the expense of more computational time.
On the contrary, when plenty of frames are at hand, it can be a good choice to limit the analysis to translation models, since the pair $ \mathrm{AUC}- \mathrm{TPR}_{@0.01} $ reports highly acceptable and comparable results with the first solution.

\subsection{State-of-the-art Comparison}\label{subsec:sota}
To the best of our knowledge, only few methods have been proposed in the literature to deal with camera attribution from stabilized videos.

One solution has been presented in \cite{taspinar2016}.
The authors only consider videos stabilized by means of a controlled algorithm (i.e., FFMPEG deshaker), which only applies rotations and translations.
As the proposed method does not take scaling into account and does not deal with videos directly stabilized on the recording device, it is likely going to fail on the uncontrolled dataset used in this paper.

A more recent solution has been proposed in \cite{iuliani2017}.
The authors propose to search for scaling and translations, but they do not take rotation into account.
Moreover, they only attribute stabilized videos to cameras if a reference PRNU obtained from still images is available (i.e., they do not compare videos to videos).
This makes their problem formulation more similar to the one we described in Section~\ref{sec:ki_v_est}, rather than the method proposed in Section~\ref{sec:kv_est}.

Additionally, both solutions proposed in \cite{taspinar2016,iuliani2017} make use of the first frame of each video sequence, which we do not consider as there is a high change it has not been stabilized, thus making the problem less challenging.

In the light of these considerations, even the comparison against \cite{iuliani2017} would not been completely fair.
However, the used metrics are the same (i.e., $ \mathrm{TPR_{@0.01}} $ and  $ \mathrm{AUC}$), and concerning the dataset, we both consider videos from the Vision dataset (8 devices in \cite{iuliani2017}, 14 devices in this paper).
Therefore, a few conclusions can still be drawn.
To compare the methods in the same experimental set-up, we select from Vision dataset all the available instances we can find for each device model exploited in \cite{iuliani2017}.
As a matter of fact, exploiting the video fingerprint $\Kiv$ described in Section~\ref{sec:ki_v_est} over this reduced dataset, we are able to achieve $ \mathrm{TPR}_{@0.01}= 0.89$ and $\mathrm{AUC}=0.96 $ using the complete strategy over $5$ query frames, and $ \mathrm{TPR}_{@0.01}= 0.92$ and $\mathrm{AUC}=0.97 $ testing $50$ query frames with the quick method.
Conversely, results in \cite{iuliani2017} show $ \mathrm{TPR}_{@0.01}= 0.87$ and $\mathrm{AUC}=0.95 $, which are below the ones achieved by us, even considering that we are discarding the contribution from the first frame, and we also cope with the video vs. video case.

\subsection{Upper bound on video fingerprint estimation}
In order to understand whether it is possible to extract better fingerprint information from stabilized video frames, we perform a final experiment involving an \emph{Oracle} providing us with data normally unavailable to the analyst.

Specifically, let us consider the scenario depicted in Section~\ref{sec:kv_est}, in which the video fingerprint $\Kv$ is extracted from video frames.
However, we envision an \emph{Oracle} telling us how to align each frame noise residual with the others, in order to obtain a much better video fingerprint estimate.
To do this practically, we apply an analogous algorithm to the one proposed in Section~\ref{sec:kv_est}, with the difference that frame alignment step is performed by similarity transformation against the reference $ \Kiv $ (i.e., a cleaner version of the device fingerprint) rather than a reference video frame. 

Of course, this is clearly an unrealistic situation (i.e., if the analyst had $ \Kiv $, he/she could use the algorithm proposed in Section~\ref{sec:ki_v_est}).
Nonetheless, this is a very powerful investigation tool for evaluating the accuracy of our results.
We can therefore compare results obtained with this \emph{Oracle}-based fingerprint, and with the proposed fingerprint $\Kv$, to see how much they differ.

In terms of quality of the estimated fingerprint, the final values of $ {\rho}(F) $ evaluated with the \emph{Oracle}-based fingerprint only report a slight increment (less than $0.1$ on average) with respect to those obtained in Fig.~\ref{fig:NCC_end} using $\Kv$. This confirms that the proposed method is quite good and represents a viable solution for extracting the device fingerprint in video domain.

\begin{figure}[t]
	\centering
	\includegraphics[width=.65\columnwidth]{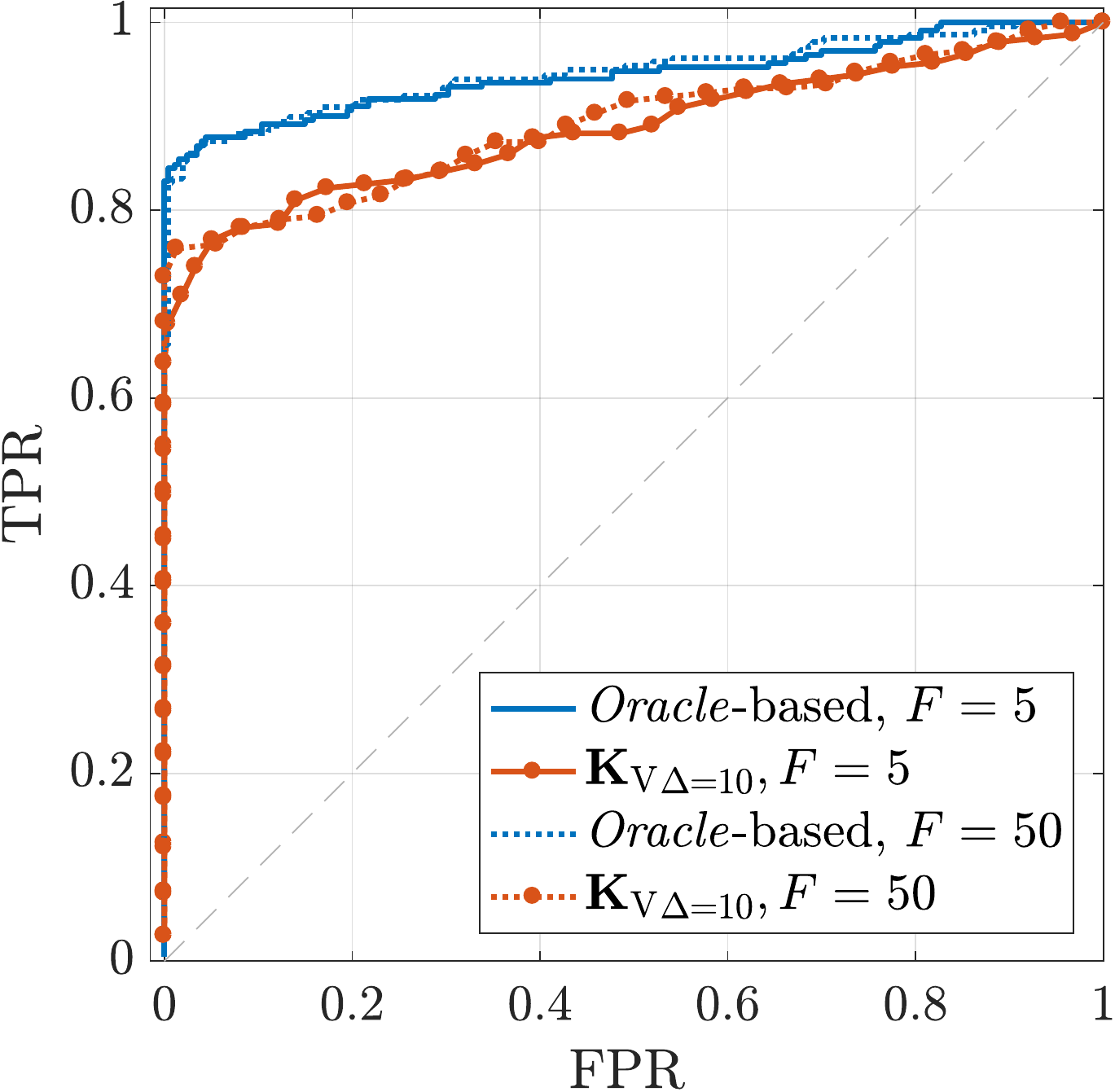}
	\caption{ROC curves obtained testing the complete strategy on $5$ I-frames, and testing the quick strategy on $50$ I-frames, using the \emph{Oracle}-based fingerprint and $\Kv$. }
	\label{fig:ROC_oracle}
\end{figure}

In terms of device attribution, Fig.~\ref{fig:ROC_oracle} reports ROC results obtained with the use of the \emph{Oracle} and results obtained with $\Kv$, using either the complete or quick test methods.
Notice that the performances do not significantly drop, on the contrary, the proposed method is reasonably accurate considering that it works in the realistic scenario where video sequences can contain some textures, potentially undermining the video fingerprint estimation.

\section{Conclusions}\label{sec:conclusions}

In this paper we propose a solution to the problem of video source attribution when motion stabilized video sequences are considered.
We devise two different solutions for estimating the camera reference fingerprint, and we also propose a technique (and a simplified version of it) to perform an attribution test between a stabilized video and a fingerprint.
The experimental campaign is conducted on a publicly available dataset composed by almost 400 videos coming from stabilized and non-stabilized mobile devices.
Notice that video stabilization is performed directly onboard by proprietary software, and we have no controls over it, thus making the experiments completely realistic.

The achieved results highlight a series of interesting aspects.
Indeed, we confirm that using the standard PRNU-based pipeline for video attribution leads to poor results.
However, with the proposed approach, it is possible to solve the problem by iteratively compensating the effect of video stabilization, even using videos only.
As a matter of fact, the best results are obtained if images are available for reference fingerprint estimation.
Using videos only, the achieved performance worsen as expected, yet not being too far from the case in which an \emph{Oracle} tells us how to correctly compensate each frame.

Another interesting aspect is that modeling video stabilization with similarity transformation proves to be quite effective.
As a matter of fact, proprietary video stabilization algorithm used by the devices under analysis are not publicly disclosed.
However, we manage to attribute videos to devices even lacking this knowledge.

Additionally, we highlight the effect of using the first acquired video frame for camera attribution.
As it is often not stabilized, considering it within the experimental campaign can produce misleading results and leads to wrong conclusions.
This is especially true if we consider a future scenario in which mobile devices will start recording videos even before pressing the rec button (i.e., as already proposed in the latest Android-based Google devices).
Indeed, in this situation, the concept of first acquired frame becomes fuzzy, and possibly all available frames can be motion compensated.

\balance

\ifCLASSOPTIONcaptionsoff
  \newpage
\fi

\bibliographystyle{IEEEtran}
\bibliography{bibliography}

\end{document}